# Architecting Safe Automated Driving with Legacy Platforms

Naveen Mohan

Licentiate Thesis
KTH Royal Institute of Technology
Industrial Engineering and Management
Department of Machine Design
SE-1044 Stockholm, Sweden

Academic thesis, which with the approval of the Royal Institute of Technology, will be presented for public review in fulfilment of the requirements for a Licentiate of Technology in Machine Design on 2018-03-08 at 10:00 in Room Gladan, Brinellvägen 85, 11428, Stockholm, Sweden.

## Abstract

Modern vehicles have electrical architectures whose complexity grows year after year due to feature growth corresponding to customer expectations. The latest of the expectations, automation of the dynamic driving task however, is poised to bring about some of the largest changes seen so far. In one fell swoop, not only does required functionality for automated driving drastically increase the system complexity, it also removes the fall-back of the human driver who is usually relied upon to handle unanticipated failures after the fact. The need to architect thus requires a greater rigour than ever before, to maintain the level of safety that has been associated with the automotive industry.

The work that is part of this thesis has been conducted, in close collaboration with our industrial partner Scania CV AB, within the Vinnova FFI funded project ARCHER. This thesis aims to provide a methodology for architecting during the concept phase of development, using industrial practices and principles including those from safety standards such as ISO 26262. The main contributions of the thesis are in two areas. The first area i.e. Part A contributes, (i) an analysis of the challenges of architecting automated driving, and serves as a motivation for the approach taken in the rest of this thesis, i.e. Part B where the contributions include, (ii) a definition of a viewpoint for functional safety according to the definitions of ISO 42010, (iii) a method to systematically extract information from legacy components and (iv) a process to use legacy information and architect in the presence of uncertainty to provide a work product, the Preliminary Architectural Assumptions (PAA), as required by ISO 26262. The contributions of Part B together comprise a methodology to architect the PAA.

A significant challenge in working with the industry is finding the right fit between idealized principles and practical utility. The methodology in Part B has been judged fit for purpose by different parts of the organization at Scania and multiple case studies have been conducted to assess its usefulness in collaboration with senior architects. The methodology was found to be conducive in both, generating the PAA of a quality that was deemed suitable to the organization and, to find inadequacies in the architecture that had not been found earlier using the previous non-systematic methods. The benefits have led to a commissioning of a prototype tool to support the methodology that has begun to be used in projects related to automation at Scania. The methodology will be refined as the projects progress towards completion using the experiences gained.

A further impact of the work is seen in two patent filings that have originated from work on the case studies in Part B. Emanating from needs discovered during the application of the methods, these filed patents (with no prior publications) outline the future directions of research into reference architectures augmented with safety policies, that are safe in the presence of detectable faults and failures. To aid verification of these ideas, work has begun on identifying critical scenarios and their elements in automated driving, and a flexible simulation platform is being designed and developed at KTH to test the chosen critical scenarios.

## Sammanfattning

Efterfrågan på nya funktioner leder till en ständigt ökande komplexitet i moderna fordon, speciellt i de inbyggda datorsystemen. Införande av autonoma fordon utgör inte bara det mest aktuella exemplet på detta, utan medför också en av de största förändringar som fordonsbranschen sett. Föraren, som ”back-up” för att hantera oväntade situationer och fel, finns inte längre där vid höggradig automation, och motsvarande funktioner måste realiseras i de inbyggda system vilket ger en drastisk komplexitetsökning. Detta ställer systemarkitekter för stora utmaningar för att se till att nuvarande nivå av funktionssäkerhet bibehålls.

Detta forskningsarbete har utförts i nära samarbete med Scania CV AB i det Vinnova (FFI)-finansierade projektet ARCHER. Denna licentiatavhandling har som mål att ta fram en metodik för konceptutveckling av arkitekturer, förankrat i industriell praxis och principer, omfattande bl.a. de som beskrivs i funktionssäkerhetsstandards som ISO 26262.

Avhandlingen presenterar resultat inom två områden. Det första området, del A, redovisar, (i) en analys av utmaningar inom arkitekturutveckling för autonoma fordon, vilket också ger en motivering för resterande del av avhandlingen. Det andra området, del B, redovisar, (ii) en definition av en ”perspektivmodell” (en s.k. ”viewpoint” enligt ISO 42010) för funktionssäkerhet, (iii) en metod för att systematiskt utvinna information från existerande komponenter, och (iv) en process som tar fram en arbetsprodukt för ISO 26262 – Preliminära Arkitektur-Antaganden (PAA). Denna process använder sig av information från existerande komponenter – resultat (iii) och förenklar hantering av avsaknad/osäker information under arkitekturarbetet. Resultaten från del B utgör tillsammans en metodik för att ta fram en PAA.

En utmaning i forskning är att finna en balans mellan idealisering och praktisk tillämpbarhet. Metodiken i del B har utvärderats i flertalet industriella fallstudier på Scania i samverkan med seniora arkitekter från industrin, och har av dessa bedömts som relevant och praktiskt tillämpningsbar. Erfarenheterna visar att metodiken stödjer framtagandet av PAA’s av lämplig kvalitet och ger ett systematiskt sätt att hantera osäkerhet under arkitekturutvecklingen. Specifikt så gjorde metoden det möjligt att identifiera komponent-felmoder där arkitekturen inte var tillräcklig för åstadkomma önskad riskreducering, begränsningar som inte hade upptäckts med tidigare metoder. Ett prototypverktyg för att stödja metodiken har utvecklats och börjat användas på Scania i projekt relaterade till autonoma fordon. Metodiken kommer sannolikt att kunna förfinas ytterligare när dessa projekt går mot sitt slut och mer erfarenheter finns tillgängliga.

Arbetet i del B har vidare lett till två patentansökningar avseende koncept som framkommit genom fallstudierna. Dessa koncept relaterar till referensarkitekturer som utökats med policies för personsäkerhet (Eng. ”safety”) för att hantera detekterbara felfall, och pekar ut en riktning för framtida forskning. För att stödja verifiering av dessa koncept har arbete inletts för att identifiera kritiska scenarios för autonom körning. En flexibel simuleringsplattform håller också på att designas för att kunna testa kritiska scenarios.

## List of appended publications:

### *Paper A:*

Naveen Mohan, Martin Törngren, Viacheslav Izosimov, Viktor Kaznov, Per Roos, Johan Svahn, Joakim Gustavsson, Damir Nesic. "Challenges in Architecting Fully automated driving; with an emphasis on Heavy Commercial Vehicles," 2016 Workshop on Automotive Systems/Software Architectures (WASA), Venice, 2016, pp. 2-9. doi: 10.1109/WASA.2016.10

Naveen and Martin have contributed to Sections I, II, III, IVE and V primarily, and across the paper in general. Naveen organized and integrated the work from the rest of the authors. Fredrik Asplund has contributed to IVB and IVD, Viacheslav has contributed to sections II and IVC. Viktor has contributed to sections II and IVA. Per, Johan and Damir have contributed to IVA. Joakim has contributed to section IVD. Every author has been part of extensive discussions through the process of writing.

### *Paper B :*

Naveen Mohan, Martin Törngren and Sagar Behere, "A Method towards the Systematic Architecting of Functionally Safe automated driving- Leveraging Diagnostic Specifications for FSC design," SAE Technical Paper 2017-01-0056, 2017, doi: 10.4271/2017-01-0056.

Naveen developed the idea and wrote the paper, Martin and Sagar provided feedback and comments.

### *Paper C:*

Naveen Mohan, Per Roos, Johan Svahn, Martin Törngren and Sagar Behere, "ATRIUM — Architecting under uncertainty: For ISO 26262 compliance," 2017 Annual IEEE International Systems Conference (SysCon), Montreal, QC, 2017, pp. 1-8. doi: 10.1109/SYSCON.2017.7934819.

Naveen developed the idea with Johan and Per. Naveen wrote the paper primarily, with contributions in specific sections from Johan and Per. Martin and Sagar provided feedback and comments.

## List of additional publications:

J. Oscarsson, M. Stolz-Sundnes, N. Mohan and V. Izosimov, "Applying systems-theoretic process analysis in the context of cooperative driving," 2016, 11th IEEE Symposium on Industrial Embedded Systems (SIES), Krakow, 2016, pp. 1-5. doi: 10.1109/SIES.2016.7509433

X. Zhang, N. Mohan, M. Torngren, J. Axelsson and D. J. Chen, "Architecture exploration for distributed embedded systems: a gap analysis in automotive domain," 2017 12th IEEE International Symposium on Industrial Embedded Systems (SIES), Toulouse, 2017, pp. 1-10. doi: 10.1109/SIES.2017.7993377

N. Mohan, P. Roos & J. Svahn, “System and Method for Controlling a Motor Vehicle to Drive Autonomously”, Patent application. Filed at Swedish Patent Office, PRV; no 1751581-8, 2017.

N. Mohan, P. Roos & J. Svahn, “System and Method for Controlling a Motor Vehicle to Drive Autonomously”, Patent application. Filed at Swedish Patent Office, PRV; no. 1751580-0, 2017.

# Acknowledgements

I would like to thank my main supervisor Martin Törngren, for being Martin. You shine through each time, particularly so in the endgame, to pull everything together and save my life! Without your advice and help in meeting deadlines, this thesis would not have been possible! I would also like to thank Sagar Behere, my co supervisor for his support and encouragement. The discussions with you have greatly helped shape the thesis.

At Scania, Per Roos and Johan Svahn have been my primary collaborators. Thank you for all the input you provided, the hours you spent with me, and (almost) more importantly for your humour. If not for your humour, discussions about deaths caused by driver distraction, would have been much more morbid.

My fellow PhD students at ARCHER/ASSUME, Joakim Gustavsson, Damir Nešić, and Masoumeh Parseh, deserve thanks for providing me with a nice working environment, and tolerating my sense of humour.

Viktor Kaznov, Sofia Cassel, Fredrik Asplund, Xinhai Zhang, Viacheslav Izosimov and Lars Svensson, thank you for the discussions and our collaborations.

I would also like to thank my colleagues within the Mechatronics and Embedded Control division, and the PhD students within Machine Design, for all the discussions we have had over lunch, fika and out of work. A special mention goes to Björn Finér, who sacrificed his weekend to fix my computer and saved about a year of work!

Several other people have been instrumental to this thesis in indirect ways. Martin Grimheden and Mattias Nyberg, during my recruitment process, were kind enough to give me time to finish tasks from my previous job. Cecilia Ollfors, Mikael Arnelind and Roger Hendelberg from QRTECH A.B., gave me the stability I needed in life, by granting me a leave of absence to pursue my doctoral studies. Thanks to the difference made by you guys, I could have a relatively smooth transition back to student life. You have my gratitude.

This thesis is dedicated to the three most important people in my life, my parents and my wife. My parents gave me life and made it meaningful. It is only due to your constant, relentless, (and many a time, undeserved) love and encouragement, that I have ever been able to achieve anything in life.

Harini, as I approach my title of half-baked PhD with this thesis, I have gone one step ahead in our arms-race towards the doctoral degree. But losing to you will be a joy of its own, catch up! Thank you for tolerating all the late nights with grace, and for taking care of every last detail when I needed it.

## Glossary

The terms used in this thesis find slightly varying usage in the different fields that are touched upon. Commonly used terms are introduced in this section as a starting point.

**ABAS: Attribute-based architectural styles** [1]**.** An architectural style in which constraints focus on component types and patterns of interaction that are relevant to quality attributes. [2]

**ADI: Autonomous Driving Intelligence.** A term from Paper A used to refer to the additions to the legacy platform required to enable automated driving, roughly corresponding to the role a human driver plays in a non-automated vehicle using the 'Observe, Orient, Decide and Act' loop [3] as a reference. Used to distinguish the additions to the platform, from the entire functionality for automation which is represented by the related term ADS.

**ADS: Automated Driving System** [4]**.** The hardware and software that are collectively capable of performing the entire DDT on a sustained basis. A term used to describe a L3, L4 or L5 system.

**AEB: Advanced Emergency Brake** [5]**.** A system that can automatically detect an emergency situation and activate the braking system, to decelerate the vehicle with the purpose of avoiding or mitigating the effects of a collision.

**Architect:** Person who performs the process of Architecting. In this thesis, an Architect refers only to those who work on the vehicular platform as a whole and is a role used within the methodology presented.

**Architecting** [6]**:** Process of conceiving, defining, expressing, documenting, communicating, certifying proper implementation of, maintaining and improving an architecture throughout a system's life cycle.

**Architecture** [6]**:** Fundamental concepts or properties of a system in its environment embodied in its elements, relationships, and in the principles of its design and evolution.

**ASIL: Automotive Safety Integrity Level** [7]**.** One of four levels (A to D) to specify the item's or element's necessary requirements of ISO 26262 and safety measures to apply for avoiding an unreasonable residual risk, with D representing the most stringent and A the least stringent level

**ATRIUM: ArchiTectural RefInement using Uncertainty Management.** Process introduced in Paper C, that is part of the methodology presented in this thesis to construct the PAA.

**Automated:** As an adjective, this term is used to denote the capability of a system, qualified in that, it does not require a human to operate it under specific conditions.

**Autonomy or autonomous**: Terms that are currently deprecated by the second edition of SAE J3016 [4]. In this thesis, they are synonymous to the term automated.

**Component:** Used synonymously with Element in this thesis.

**Concern** [6]**:** Interest in a system relevant to one or more of its stakeholders.

**DDT: Dynamic Driving Task** [4]**.** All the real-time operational and tactical functions required to operate a vehicle in on-road traffic, excluding the strategic functions such as trip scheduling and selection of destinations and waypoints**.**

**Design decisions:** Within these thesis, design decisions refer only to decisions made in the process of architecting, in the selection of one or more choices that influence the system under consideration.

**Diagnostic monitor:** A test run by ECU software to detect the presence of faults. When enabled according to its criteria, and if configured to, it may save a DTC and optionally some snapshot information.

**Diagnostic specification:** For the purposes of this thesis, a specification of the system that contains linked information about detectable faults, the monitors that detect them, the DTCs that are set and the functional effects of the faults.

**Domain expert:** A role defined in the methodology presented in this thesis, denoting experts in specific areas or technologies.

**Driver or human driver:** Human who is performing or will perform the DDT.

**DSL: Domain Specific Language.** A modelling language specialized for a specific domain.

**DTC: Diagnostic Trouble Code**. An identifier saved by an ECU, in case of a detected fault by one or more monitors.

**E/E systems: Electrical/Electronic systems** [7]**.** System that consists of electrical and/or electronic elements including programmable electronic elements

**EBS: Electronic Braking Subsystem**. For the purposes of this thesis, the subsystem for Service brakes at Scania C.V.

**ECU: Electronic Control Unit.** A term for an embedded system that provides some functionality, typically to control a E/E subsystem, in a vehicle.

**Element:** A system or part of a system including physical parts such as ECUs, sensors, and actuators; and functional components which can be individual software units as well as the logical aggregation of these components into functional elements.

**Extra-functional requirement:** A requirement representing the qualities or properties of the function or services that the system provides. In this thesis, these are typically related to either safety or reliability.

**FSC: Functional safety concept** [7]**.** Specification of the functional safety requirements, with associated information, their allocation to architectural elements, and their interaction necessary to achieve the safety goals.

**FSR**: **Functional Safety Requirement** [7]**.** Specification of implementation-independent safety behaviour, or implementation-independent safety measure, including its safety-related attributes

**Functional requirement:** Requirement that specifies the services or functions of the system.

**Functional Safety** [7]**:** Absence of unreasonable risk due to hazards caused by malfunctioning behaviour of E/E systems.

**Or**

**Functional Safety:** A viewpoint defined in Section 3.4 of this thesis.

**HGV: Heavy Goods Vehicles.** Vehicles that have a total weight of above 3500 kg. These are typically used for commercial purposes.

**Item** [7]**:** System or array of systems to implement a function at the vehicle level, to which ISO 26262 is applied.

**Legacy:** In this thesis, this term is used as an adjective to denote the availability of elements at the start of the application of the process or method.

**L*x*; where x is 0 to 5:** automated driving functions of level x as per the SAE system of classification for driving automation [4].

**Method** [8]**:** Consists of techniques for performing a task, in other words, it defines the "HOW" of each task.

**Methodology** [8]**:** A collection of related processes, methods, and tools.

**Minimal Risk Condition:** A condition to which a user or an ADS may bring a vehicle to reduce the risk of a crash when a given trip cannot or should not be completed [4].

**OEDR: Object and Event Detection and Response** [4]**.** The task of monitoring the driving environment and executing a response.

**OEM: Original Equipment Manufacturer**. In the context of this thesis, represents automotive manufacturers who series-produce complete vehicles that are sold to customers.

**PA: Preliminary Architecture.** A term originating from ISO 26262, redefined in this thesis to be a set of functional elements that together represent the item.

**PAA: Preliminary Architectural Assumptions.** A term from originating from ISO 26262 refined in Section 3.4 of this thesis, as a view of the architecture governed by the Functional Safety viewpoint.

**Perceived Certain Domain:** That part of the information domain that contains known information, as well as information that is assumed, for the purposes of making an analysis. Introduced in Paper C.

**Platform:** A platform is the set of all elements that provide Electrical/Electronic (E/E) functionality for a vehicle including (but not limited to) software, hardware and associated low-level control.

**Process** [8]**:** A "Process (P) is a logical sequence of tasks performed to achieve a particular objective. A process defines "WHAT" is to be done, without specifying "HOW" each task is performed." At any level, process tasks are performed using methods.

**Risk:** Within this thesis, risks refer to project risks typically related to safety caused by design decisions made during architecting.

**Safety case** [7]**:** Argument that the safety requirements for an item are complete, and satisfied by evidence compiled from work products of the safety activities during development

**Safety engineers:** A role defined in the methodology presented in this thesis, denoting engineers who are tasked with assuring functional safety of the product.

**Safety goal** [7]**:** Top-level safety requirement as a result of the hazard analysis and risk assessment

**Safety:** Used synonymously with Functional Safety within this thesis

**Series production**: The manufacturing of goods in large quantities using standardized designs and processes.

**Service brakes:** The braking system that uses friction on the wheels as a way to retard the motion of the vehicle.

**Snapshot:** Data saved by a monitor when it reports a DTC, for the purposes of debugging or recording system state. Typically contains information about the conditions that triggered the fault and values of variables related to it.

**Software-intensive systems** [6]**:** A system where software contributes essential influences to the design, construction, deployment, and evolution of the system as a whole.

**SOTIF** [9]**:** Safety of the Intended Functionality. Currently a draft standard (ISO/WD PAS 21448 Road vehicles – Safety of the intended functionality) intended to address safety in the case of hazards that are not related to functional failures.

**Subsystem:** A set of related functional Elements.

**System safety:** Absence of unreasonable risk. Note that this is not limited to errors caused by malfunctioning systems as is the term Functional safety.

**Tool** [8] : an instrument that, when applied to a particular method, can enhance the efficiency of the task; provided it is applied properly and by somebody with proper skills and training.

**UML: Unified Modelling Language**. A graphical modelling language for visualizing, specifying, constructing, and documenting artefacts of a software-intensive system.

**Uncertain domain:** The part of the information domain that contains information that is subject to the possibility of change with new knowledge. Introduced in Paper C.

**View or architecture view** [6]**:** Work product expressing the architecture of a system from the perspective of specific system concerns.

**Viewpoint or architecture viewpoint** [6]**:** Work product establishing the conventions for the construction, interpretation and use of architecture views to frame specific system concerns.

# Contents

## List of Figures

## List of Tables

# 1. Introduction

The concept of the colloquially known autonomous driving, or the practical automation of the dynamic driving task (DDT) [4], has been around for many years now. As early as the 1960s, there have been visions of a future where cars could drive themselves and the drivers could be passengers instead using the time saved for more productive purposes. In the mid-90s researchers at Carnegie Mellon University claimed to have driven over 6000 steering miles autonomously [10], and similar claims were made by researchers from Universitat der Bundeswehr München [11]. However, as the reality of the complexity of the task emerged, many of these ideas faded away and automated driving remained seemingly out of reach. Though there have been pushes in the direction of automated driving in the past, a major change that has facilitated the most recent one is the increase in the performance of sensory technologies and computing power allowing development of algorithms to enable machine inference in real-time. While there have been many prototypes of automated vehicles demonstrated both in academia [12] and in industry [13] [14], these are typically still in various levels of development and as of late 2017, no OEM has a highly or fully automated on-road vehicle that has been series produced.

To reach the goal of profitable series-production, the automated functions must be integrated within the complex vehicular architectures most OEMs use today. Modern vehicular architectures differ greatly from their predecessors from about a couple of decades ago where only a few components, if any, were controlled electronically. These individually controlled embedded systems have since transformed into progressively larger interconnected distributed systems, to the point that a modern vehicle may have greater than 100 separate Electronic Control Units (ECUs) providing close to about 300 functions of value to the user. In the cost sensitive automotive domain, the maintenance of these functions, hardware and software components, into structured product lines i.e. creating a platform of reusable components, brings an additional dimension of complexity.

Platform based development, cost, functional safety, and their influence on the process of architecting automated driving in Heavy Commercial Vehicles (HGVs), were some of the main drivers behind the Swedish national research (Vinnova-FFI) project ARCHER [15]. ARCHER was started as a joint effort between KTH and Scania CV AB and has been the main source of funding for this work. The goals of the project are as follows:

> *"The overall goal for this project is to develop methods and principles and a reference system architecture enabling safe, secure and cost-efficient fully automated heavy vehicles. The "deliverable" is further described as a system architecture description that incorporates an architectural realization of the safety concepts, testability mechanisms, necessary to show the feasibility of building a fully automated heavy vehicle, exhibiting sufficient robustness to provide realistic availability of operation in the considered operational scenario. The reference system architecture has the goal to improve on existing prototype systems for various levels of* *automated driving by, 1) not assuming any operator presence, and, 2) being feasible for series production of commercial vehicles.” [15]*

This thesis, as a part of ARCHER, focusses on methodologies for architecting in the context of automated driving, functional safety, and industrialization concerns such as cost and reuse of legacy. Figure 1 shows a high-level view of the scope. As an overarching idea, this thesis views automation to be an add-on to an existing platform of components, aims to identify the challenges in integrating automation into the platform in a functionally safe way, and to define methods, processes, and tools to ameliorate these challenges. To this end, while the work presented in the papers is often performed in the context of automated driving and heavily utilizes the core principles of automated driving and the technologies involved, the focus remains squarely on the ***methodologies for architecting***. The methodology presented in this thesis is designed for industrial usage and defines the architectural viewpoint [6] of Functional Safety to create the view of Preliminary Architectural Assumptions (PAA) complying with the current dominant functional safety standard ISO 26262 [7].

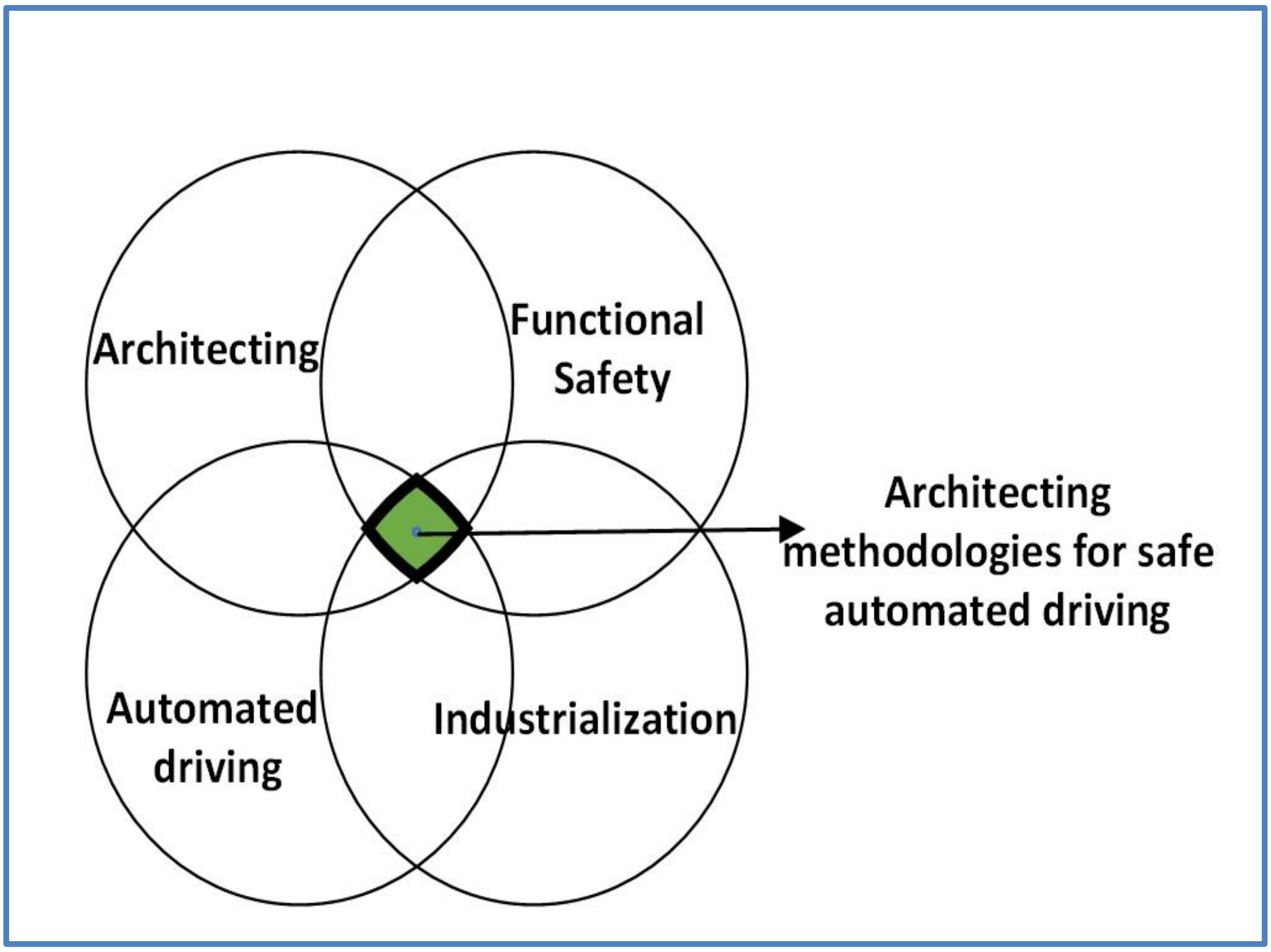

**Figure 1: Focus of this Thesis**

## 1.1 Assumptions and Delimitations

Due to the rather broad range of the topics involved, several delimitations had to be made on the scope of this work. It is useful to take a pause here to clarify what assumptions were made and what the thesis will *not* be about.

This work relates only to ***on-road*** vehicles and thus, does not take into consideration other automated vehicles that work in more constrained environments, such as Autonomous Guided Vehicles in factories, automated trains etc.

A delimitation of this work is that it is intended to be applied to the concept phase in the development lifecycle and looks at architecture from a ***functional perspective***. Thus, implementation or technology specific considerations such as the choice of processor specific architectures, which are choices made in the latter phases of development, are out of scope. There are assumptions of a middleware being used (as is common in the industry see e.g. AUTOSAR [16]) that will be able to isolate the functionality of automation from the hardware implementation details as is common in the industry, but a specific ***choice of middleware is not made***. While considering legacy systems such as in Paper B or judging the technological options available for a use-case and feasibility of implementation, it is sometimes unavoidable to investigate the details of the technology itself. Paper C touches on this topic as well from the abstraction of architectural decisions. However, these technological points can get very detailed (see for example - [17] for considerations on radar technologies) and care has been taken to keep the ideas behind paper B and C ***independent of technology*** specifics as far as possible.

Safety, particularly ***functional safety***, is the primary extra functional concern of this thesis and others such as security of automated systems, though considered for e.g. in the case studies are not the focus of this work. Issues regarding functional safety that are not related to architecting in the concept phase e.g. tool-qualification etc. are not specifically investigated. The focus remained on faults and safety measures, related to hazardous events and the ***movement of the vehicle*** - for example, collisions and not fires caused due to burnt fuses. A similar delimitation on the evaluation of completeness or correctness of requirements, in that they are assumed to be correct for the purposes of this thesis.

In terms of architecting itself, an assumption is that a ***functional view of the legacy platform*** architecture is available in that it describes a set of functional components and the interactions between them. Detailed issues regarding the modelling of the legacy components, such as architecture recovery, are not considered within this thesis. The focus remains on the ***bigger picture***. Similarly, the variability of these components is not explicitly considered within this work. Automated driving is assumed to be introduced to the platform in ***stages*** that bring in progressively higher levels of automation to progressively larger parts of the platform.

Some details regarding automated driving such as the ethics, morality, legality, and social desirability of automated driving are left out as research topics of their own. The infamous trolley problem etc. are thus out of scope and not part of this thesis. The case studies primarily drew the system boundary at a ***single truck*** and do not include the aspects of intelligent transport systems, or even platooning, eschewing the *system of systems* view of thinking and avoiding the dependencies on communication between vehicles.

## 1.2 Thesis Outline and Reading Guide

Drawing from ISO 42010 [6], this thesis defines the PAA as a view of the architecture governed by the concern of functional safety and delivers a methodology for architecting the PAA.

This thesis is organized in the following way: Section 2 gives an overview of terms and challenges to set the context for the specific aims of the thesis in Section 3. Section 4 then describes the research design, methods and limitations allowing section 5 to focus on the content of the paper itself. Section 5 will provide a concise summary of the included papers and a description of how the papers tie together as part of a greater whole. Section 6 concludes with a summary of the impact and the directions of future work.

Though the thesis is intended to be read in the order presented, some readers may be interested in only a subset of the topics addressed. A guide to the important parts of the thesis is as follows:

- For the basis of the research itself, Section 3.2 presents the research questions and Section 4 details its philosophical aspects, choice of methodology, threats for validity etc.
- For a primer on the different fields this thesis touches upon, Sections 1, 2 and 3.1 are important.
- For the identified challenges, refer to Sections 2.4, 3.1 and Paper A.
- For an overview of the related work, see Section 3.3.
- For the main contributions, the reader is directed to Sections 3.4 for the definition of the Functional Safety viewpoint and Section 5.4 for a discussion of how the various parts of the thesis tie together into a methodology.
- Sections 5.1, 5.2 and 5.3 provide summaries of the included papers.
- The appended papers at the end of the thesis, provide additional details about the usage and contribution

# 2. Background and Context

This section will describe the background of legacy platforms, ISO 26262, automated driving, and the role of the driver in design of the platform, setting the context for the rest of the thesis.

Human factors and errors have been known to be the overwhelming cause of accidents when judged in comparison to vehicular and environmental factors, e.g. about 93% of the total in the study in [18]. While automation essentially replaces the human driver, ideally removing the human factors of the driving task involved, it must be noted that the automation is not infallible, and the complexity of the newly added components and their interactions will add new ways for failures to occur.

The state of California in the United States mandates the reporting of disengagements i.e.

> *"A deactivation of the autonomous mode when a failure of the autonomous technology is detected or when the safe operation of the vehicle requires that the autonomous vehicle test driver disengage the autonomous mode and take immediate manual control of the vehicle. "* [19].

Waymo inc., the spinoff from Google, was arguably the best performer from the reporting of miles driven in 2017 by the various organizations who test automated driving in California with an impressive number of only 0.178 disengagements for every 1000 miles driven [20]. The impressiveness of the figure diminishes however, when the numbers are taken in context e.g. that about 3.2 trillion miles were driven in the United States in 2016 according to official statistics [21]. Not all disengagements lead to an accident, but they do indicate a potential for one, if there is no driver ready to respond. Small numbers of faults seen in testing may have drastic impact at the scales of operation on transition from a prototype to a product. This is a fundamental reason to have a systematic approach to functional safety. The following subsections will detail the current state of practice in the industry.

## 2.1 Legacy Platforms and HGVs

The automotive industry has seen modern vehicles transform slowly but steadily from a purely mechanical design to the complex software-intensive cyber physical system it is today. The modern automotive platform can have about 100 physically disparate ECUs providing about 300 functions distributed across them and it was estimated in 2006 that about 40% of the in-production costs can be attributed to software [22]. This evolution has led to modern cars being fitted with tens of millions of lines of code and large parameter sets, see e.g. [23]. A visualization of the code base of a single 2013 Scania vehicle, Figure 2, shows about 1400 points that represent the functional elements (each functional element being a part of a function) connected by about 15000 lines representing their communication dependencies, shows the scale of penetration of software. The statistics refer to a pre-automation era of automotive platforms and automation is predicted to increase the functionality and the complexity of E/E systems used in the automotive industry.

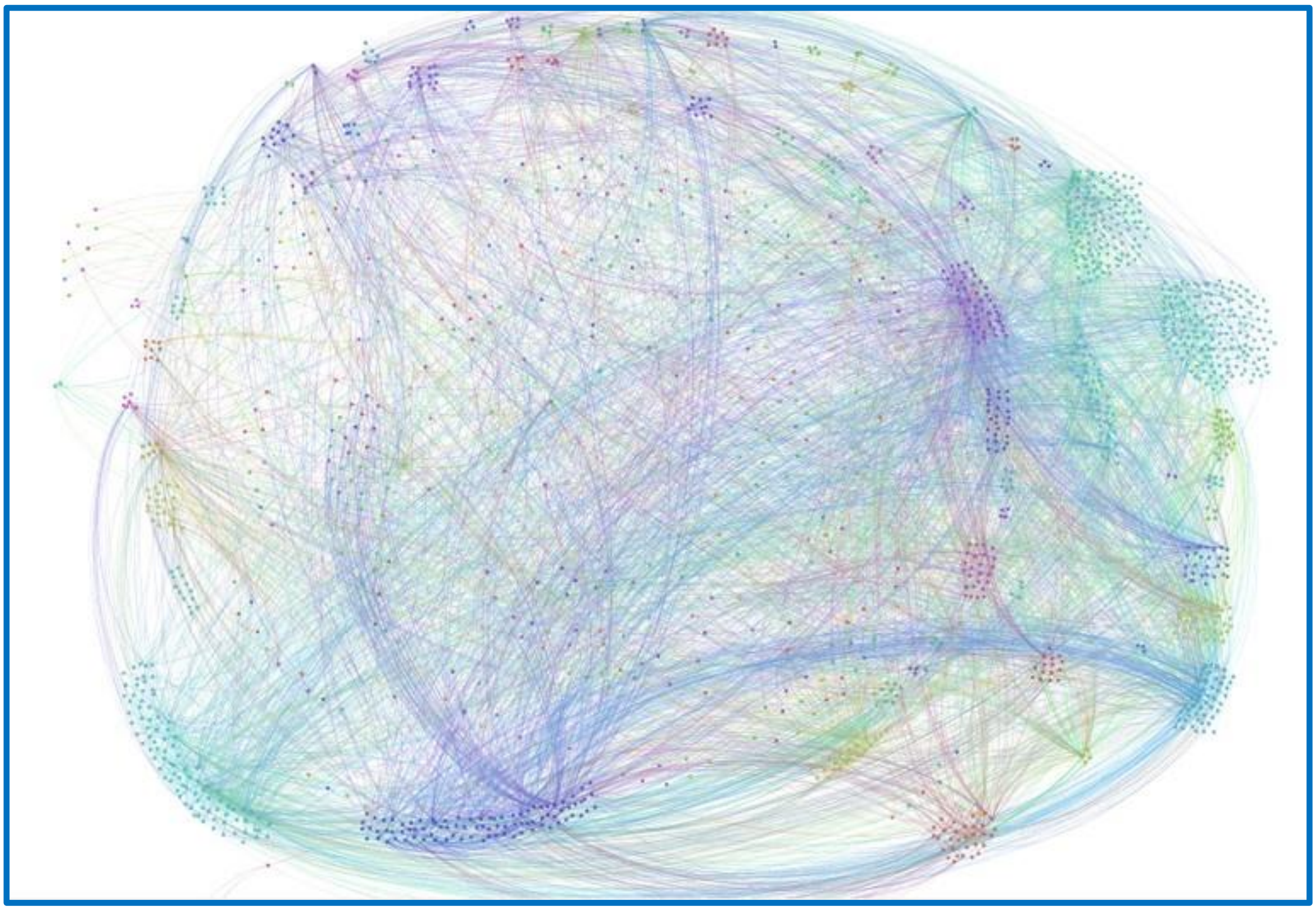

**Figure 2: Visualization of the Complexity of a Platform; Courtesy of Scania C.V.**

Modern vehicular platforms are additionally characterized by high variability caused by the customization to satisfy customer needs. Each customization or new feature adds costs which are an important concern in the automotive domain [24]. Platforms have thus evolved in the form of systematically handling product variability with product lines and families to save costs [25]. A single vehicle, such as the truck from Figure 2, is just one of the variants possible to be created from a single product line, which is in turn part of a larger product family. Suitable reading on the topic of variability may be found in [26]. A HGV is seen as a tool providing services of value to a customer and is heavily customized according to the customer’s business needs be it in hardware, software, or mechanical components. Other considerations that are unique to HGVs, such as vehicles being repurposed in the aftermarket for an entirely different purpose etc., typically add the possibility for a customer to optimize for total lifecycle costs. A larger discussion of lifecycle costs at the customer and how that affects engineering decisions during design can be found in [27]. The need to isolate common functionality for the purposes of variability and the mechatronic nature of the automotive domain, i.e. the tight coupling between the mechanical components and the E/E systems that control them, has led to a pattern of ECUs connected over busses, implementing related parts of several distributed functions [28].

Changes to a particular distributed function could influence others due to their shared hardware platform dependencies. Platform based design such as described in [29] is a core tenet in the automotive industry and architects have to optimize the architecture across a larger scope, rather than a single vehicle. Axelsson in [30] describes two types of architecting processes, revolutionary and evolutionary, which differ in the scale of their efforts. An evolutionary step is taken in the daily work of the architects when the architecture is modified to respond to smaller stimuli such as changes to requirements. A revolutionary step is a usually a planned response to a bigger stimulus and considers the platform as a whole, dimensioning it for the future. This thesis considers the introduction of automated driving to be a revolutionary change to the vehicular platforms and sees this as a motivating factor for the use of a structured methodology.

## 2.2 ISO 26262 and the Safety Lifecycle

The extra-functional property of safety has always been deeply connected to the automotive world. Recognizing the trend of the increasing complexity of E/E systems and greater dependence on software in the automotive industry, emphasis has been placed in recent years in the assurance of safety in the presence of functional failures within the system i.e. functional safety. The current de-facto standard for functional safety within the automotive domain is ISO 26262 [7], an automotive specific specialization of the “meta-standard” for safety IEC 61508 [31]. Despite known issues of using ISO 26262 such as 1) the current edition of ISO 26262 not being applicable to HGVs and 2) debate on the applicability of the standard for automated vehicles [32], the standard remains the closest “fit” and the main alternative to analyse functional safety until the publication of SOTIF [9], intended as a complement to ISO 26262, is available.

Functional safety within the standard, is treated as the absence of unreasonable risk due to hazards caused by malfunctioning behaviour of E/E systems. ISO 26262 gives a high-level reference lifecycle, and detailed processes for technical development that need to be followed, to ensure functional safety of an Item being developed. An overview of the lifecycle and processes, is seen in Figure 3.

A safety case, i.e. the argument that the safety requirements are complete and satisfied by evidence, is the desired result of applying ISO 26262 in development. Evidence is obtained by generating work products that fulfil the requirements set by the standard. The requirements are designed for the identification and mitigation of hazards related to malfunctioning behaviour of the system under consideration in the concept, product development, production, and operation phases. This thesis limits itself to the tasks in the concept phase and the following subsections summarize parts of the standard that are the most relevant to this thesis.

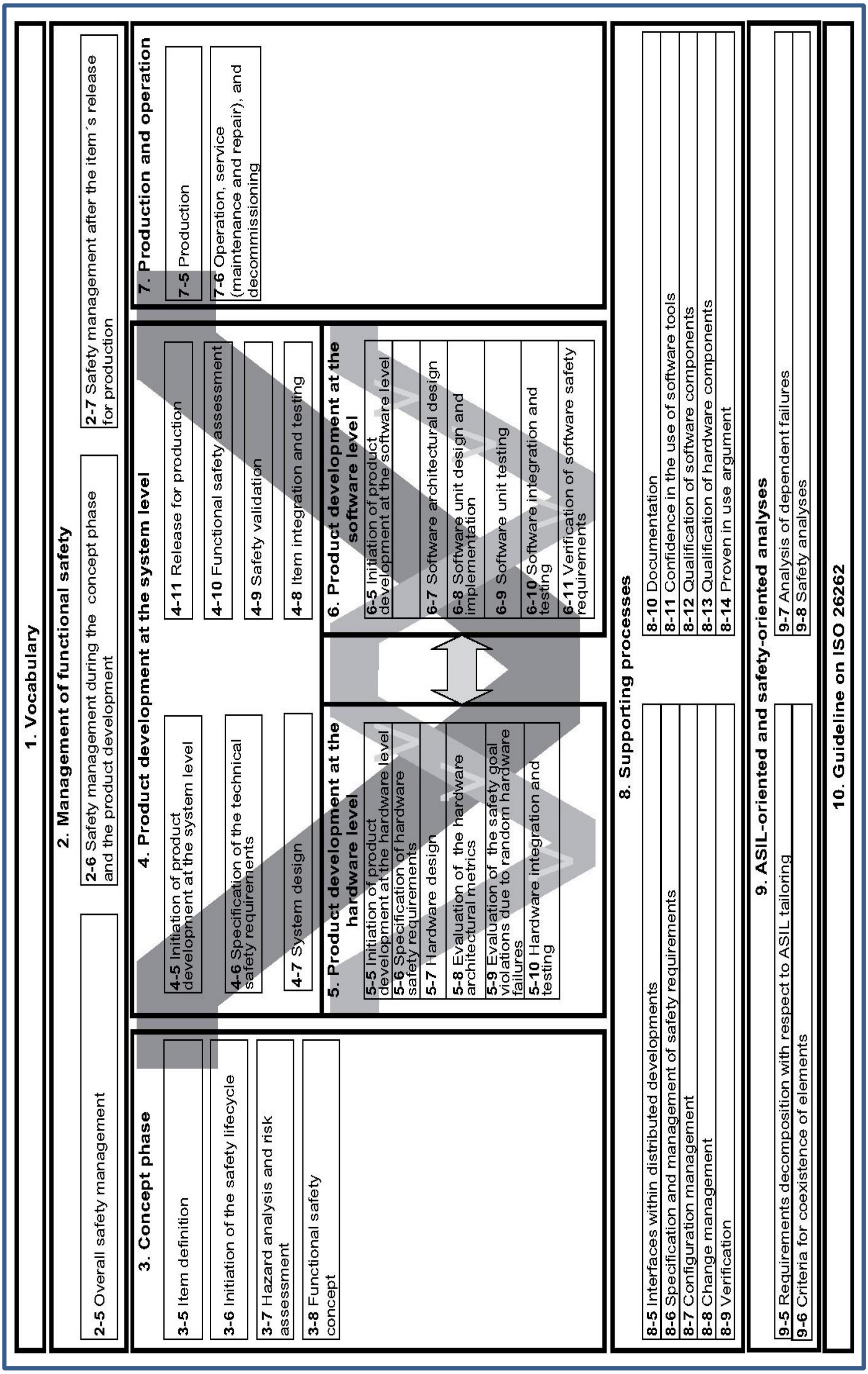

Figure 3: Overview of Processes in ISO 26262; Copyright remains with ISO

## *Hazard Analysis and Risk Assessment*

The definition of the Item is the starting point within the safety lifecycle. An Item is the system or array of systems that implements functionality at a vehicle level, serves as a demarcation of where ISO 26262 will be applied. Once the item has been sufficiently defined in terms of the requirements stated in clause 5.4.1 of the standard, the "hazard analysis and risk assessment" step is initiated to systematically identify hazards and analyse the risk presented by them. As a means of risk assessment, ASIL classification is then performed by assigning the Severity (S), Exposure (E) and the Controllability (C) of the hazard to one of their pre-determined classes. An ASIL classification is then made according to the scheme shown in Table 1, where ASIL classifications are indicated in red. A safety goal is formulated for each hazardous event that was not classified QM. Safety goals act as the top-level safety requirements for an item and all safety requirements through the lifecycle are derived from these.

**Table 1: Scheme for ASIL Classification; Adapted from ISO 26262 [7]**

| **Severity class** | **Probability class** | **Controllability class** | | |
|---|---|---|---|---|
| | | **C1** | **C2** | **C3** |
| **S1** | **E1** | QM | QM | QM |
| | **E2** | QM | QM | QM |
| | **E3** | QM | QM | A |
| | **E4** | QM | A | B |
| **S2** | **E1** | QM | QM | QM |
| | **E2** | QM | QM | A |
| | **E3** | QM | A | B |
| | **E4** | A | B | C |
| **S3** | **E1** | QM | QM | A |
| | **E2** | QM | A | B |
| | **E3** | A | B | C |
| | **E4** | B | C | D |

## *Functional Safety Concept and the Preliminary Architectural Assumptions*

The objective of the Functional Safety Concept (FSC) is to derive Functional Safety Requirements (FSRs) derived from safety goals, and allocate them to the elements of the Preliminary Architecture (PA) or to external measures ensuring that the safety goals are met. The FSC addresses the fault detection, mitigation, arbitration logics and available safety measures to argue for the safety of the concept chosen. The aim at this stage in the lifecycle is to establish implementation-independent safety behaviour to ensure that the safety goals are met. The PA is assumed, in the standard, to be delivered by available with a similar implementation-independent behaviour. ISO 26262 tends to use the terms PA and PAA interchangeably. A distinction is made in this thesis between the collection of elements of the architecture (PA), from the view governed by the Functional Safety viewpoint defined in 3.4 (PAA). The FSRs inherit the ASILs from the SGs they are derived from, and the elements the FSRs are allocated to are developed according to the rigour of that ASIL level.

### Requirement Decomposition and ASIL Tailoring

Higher ASIL requirements being allocated to elements have a direct implication to the cost of the development efforts due to the rigour associated with their development. In many cases it may be advantageous to use a “tool” provided by ISO 26262 Clause 9-5, namely the “decomposition of requirements with respect to ASIL tailoring”, more commonly known as ASIL decomposition. ASIL decomposition allows a single requirement to be broken down into two redundant ones with each having lower ASIL than the one they were decomposed from, on the basis that there are ***sufficiently independent architectural elements*** available to individually fulfil the requirements. The list of allowed ASIL decompositions is as follows:

ASIL D -> ASIL C (D) + ASIL A (D)

ASIL D -> ASIL B (D) + ASIL B (D)

ASIL C -> ASIL B (C) + ASIL A (C)

ASIL B -> ASIL A (B) + ASIL A (B)

ASIL <Any level> -> ASIL <Any level> + QM

ASIL decomposition can be applied at any stage in the development of the product and serves as an indispensable tool for an architect to address scenarios such as cases where, (1) it may be impossible for complex elements to meet the rigour needed by a high ASIL level, (2) it may be too expensive to meet the rigour needed by a high ASIL level or, (3) the element that needs to be integrated may not have been developed according to ISO 26262. If correctly used, ASIL decomposition can be used to tailor the ASIL classified requirements to match the capabilities of the elements in the item and potentially save costs. A guide to proper usage of ASIL decomposition along with common mistakes made in this process can be found in [33]. ASIL decomposition serves as an important motivating factor in the studies in Paper B, (deriving information about legacy components to identify how they should be used in a new context) and Paper C (using the information gathered to justify the predicted absence of single point failures) due to its cost saving potential.

## 2.3 Driving Automation

In popular opinion, automated driving is set to bring in a new era of accident and congestion free roads, greater urban land use with reimagined cities, better resource usage and greater productivity for everyone. According to the research and advisory firm Gartner, that assesses the maturity, adoption and the social application of new and emerging technologies, automated driving is starting to leave the top of the Gartner hype curve in 2017 [34]. This move from the top, in Gartner's methodology indicates a small step towards the maturity of the technology and more realistic expectations from investors and society. But considering that this has been a very hot-topic for the last few years and there are different views on what automated driving is, and on its nature, it is useful to establish the definition of automated driving within this work.

Automation in the purest sense of the word exists in many forms already in most modern vehicles. In most cases the automation is *limited in combinations of responsibility or duration or the Operational Design Domain (ODD).* There are for example, advanced ADAS functions that take control of the vehicle *when enabled by the user* (such as in the case of Adaptive Cruise Control functions), or active safety functions such as the AEB (legally mandated for trucks [5]) that while enabled all the time, only overrides the driver's commands *for a short time* under specific conditions. Many OEMs also have released a limited "autopilot" functionality that takes over the task of driving for sustained periods under highway conditions. A common characteristic to these functions is that the driver is ***always*** responsible. Current legal frameworks in most countries that follow the Vienna convention [35] are also built under this assumption that the driver, however assisted, is responsible for the safety of the vehicle and other road users.

There have been a few previous attempts at classifying automation such as by the German federal highway research institute (BASt) [36] and American National Highway Traffic Safety Administration NHTSA [37]; these have been deprecated in favour of the SAE levels of automation defined in SAE J3016. Table 2 shows the differences in the various terms used. Automation referred to in this thesis deals only with SAE levels 3-5 of automated driving characterized in that while the ADS is engaged, it is simultaneously responsible for:

i. complete lateral and longitudinal control of the vehicle in accordance with road rules.
ii. the monitoring of the environment for other road users and hazards
iii. the monitoring of the environment to ensure that it is still within its ODD
iv. the monitoring of vehicle performance
v. taking appropriate action to avoid hazardous events
vi. transitioning to (or attempting to transition) a minimal risk condition in case of a situation it cannot handle

**Table 2: SAE Levels of Driving Automation; Adapted from [4]**

| SAE Level | Name | *DDT* | | *DDT fallback* | *ODD* | *BaST Level* | *NHTSA Level* |
|---|---|---|---|---|---|---|---|
| | | *Sustained lateral and longitudinal vehicle motion control* | *OEDR* | | | | |
| ***Driver* performs part or all of the *DDT*** | | | | | | | |
| **0** | **No Driving Automation** | *Driver* | *Driver* | *Driver* | n/a | Driver only | 0 |
| **1** | **Driver Assistance** | *Driver* and System | *Driver* | *Driver* | Limited | Assisted | 1 |
| **2** | **Partial Driving Automation** | ***System*** | *Driver* | *Driver* | Limited | Partially automated | 2 |
| ***ADS* ("*System*") performs the entire *DDT* (while engaged*)*** | | | | | | | |
| **3** | **Conditional Driving Automation** | *System* | ***System*** | *Fallback-ready user* | Limited | Highly automated | 3 |
| **4** | **High Driving Automation** | *System* | *System* | ***System*** | Limited | Fully automated | 3/4 |
| **5** | **Full Driving Automation** | *System* | *System* | *System* | **Unlimited** | *n/a* | |

Another way of looking at the delimitation of SAE L3-5 being the focus of this thesis is that the driver is no longer involved in the supervision of the function of the ADS. Thus, the ADS must be capable of handling failures that may occur since the driver cannot be relied upon to act immediately. The significance of this choice is expanded upon in the next subsection.

## 2.4 Role of the Driver

The introduction of driving automation is driven partly by the potential to improve accident statistics; traffic statistics from the U.S.A. show that the mean time between crashes that cause injuries is about 50,000 vehicle hours [38]. The role of the driver in a vehicle will change significantly with driving automation, from operator to supervisor of the driving task, requiring taking over control only in case of situations that are beyond the scope of the ADS. The consequences of such a role change are non-trivial and parallels can be found in the domains, where automation is more mature, such as industrial processes where Bainbridge in [39] notes that reliance on automation may reduce the operator's skills. This translates to a greater potential for risk, if and when the ADS requests a takeover of the DDT by the driver. Parasuraman and Riley in [40] suggest that though automation may reduce operator error, it may instead leave the system more vulnerable to designer error. In the specific context of the automotive domain, a careful analysis of the impact of replacing the driver with the ADI is needed to minimize the errors in design. There are implications to the architecting process itself that require a reconsideration of how development is done today. Some of the ways the driver affects the development of the platform are described in the following subsections.

### *Comfort*

Traditionally, there has been emphasis on the comfort of the driver over performance or other extra-functional requirements, e.g. the frequency of operation for an electric motor may be adjusted for human comfort, packaging of elements to fit aesthetics of the cabin, cabin climate control systems, etc. With the absence of the driver in automated vehicles, many of these design choices made will eventually be revisited. This is especially true in the case of HGVs where in the common tasks of delivering goods, despite large design compromises being made for the presence of the driver, he/she is more of an unavoidable necessity than a need. The removal of the driver gives an opportunity to revisit design choices that were not available before, leading to a larger design space.

### *Complexity*

Brooks in [41] describes two different types of complexity, accidental and essential. Essential complexity is the complexity that is inherent to the application due to its nature and accidental complexity is a by-product of choices made in the design process. The ADS will need to perform at a similar level as a highly trained human driver to ensure safety (a complex task in itself) i.e. the Essential Complexity of the task but achieving it will create dependencies between the other parts of the platform and require more stringent verification than what is needed today i.e. the Accidental Complexity.

### *Minimal Risk Condition*

ISO 26262 requires that a safe state be achieved in case of a fault that could lead to a hazardous event. A safe state in the case of a fault in the ADS is addressed in the J3016 as the Minimal Risk Condition. What this means in practical terms is that the ADS will need to have redundant measures built in across most of its elements such that no single fault will stop it from achieving a Minimal Risk Condition, further adding to the complexity of the platform.

### Controllability

Controllability within ISO 26262 currently refers to the ability of the driver or others at risk, to act and mitigate the severity of a hazardous event. As seen in Table 1, a higher controllability can reduce the ASIL level assigned to a particular hazardous event. The absence of a human driver monitoring and taking immediate action in case of a hazardous event means that 1) the ADS will unavoidably have requirements with high ASIL classification and 2) legacy elements from the platform that the ADS will rely on will have to be reanalysed as they will inherit higher ASIL requirements than they were originally designed for. Thus, the complex platform will have to perform with a higher integrity which will lead to higher costs.

### Detectability vs Robustness

In a similar vein to that of Controllability discussed in the previous subsection, there has also been reliance placed on the driver's ability to detect a hazardous event. If a fault or its effects is detected by a driver in time, it becomes easier to control. E.g. a short circuit in an infotainment system may be detected immediately by software but that it has led to a fire can only be measured by a well-placed sensor in the platform. Thus, the driver's replacement affects the platform in more ways than just the absence of a controller, since the OEM will be responsible for the safety of the passengers and goods it carries, reusing legacy systems need to be reanalysed for this loss of detectability. Essentially, this means that legacy systems cannot be reused directly as off-the-shelf components and that new elements may needed to be added to the platform that have never been used before, e.g. auditory sensors to triangulate the location of an emergency vehicle.

# 3. Research Objectives

This chapter describes the approach taken in this thesis and the reasoning behind it. Section 3.1 gives brief overview of design decisions, and a problem formulation in the intermingling of architecting and safety. Section 3.2 then defines the specific research questions, followed by a discussion of the related work on the topic in Section 3.3. This section ends by with Section 3.4 defining the viewpoint that will be used in the rest of this thesis.

## 3.1 Problem Formulation

The nature of architecting limits the level of formalization possible in the design of a complex platform. There is thus a need for an "optimum" amount of rigour in the process of architecting, that adds value in the context of safety-critical systems and automated driving, while not impeding the momentum of development.

The multi-dimensional nature of architecting has been described by Rechtin in [42], as an art as much as it is a science. Architecting automotive platforms requires consideration of not only the properties of the needed function but also requires a pragmatic approach to cost, the competences in the organization, knowledge of target markets, potential future functional needs, limitations from legacy etc. As a result, the solution chosen is often a compromise between the various requirements. An architect must judge these numerous factors, often under the presence of uncertainty in the information, and make design decisions that will influence the platform for years to come. Since many of the issues trade-off against each other and are difficult to define, e.g. technology competencies in an organization, a formalization of the process of architecting remains equally difficult, and the design space prohibitively large. Architectural exploration strategies and their limitations have been covered more in detail in [43] where the authors attempt to summarize the gaps between industry and academia in the adoption of automated architectural exploration. A consequence of the lack of automated architectural exploration is that manual methods are frequently used in the automotive industry for architecting, see e.g. [44], [45], and [24], and architectural decisions are made frequently as a result of consensus amongst the stakeholders. While consensus has been ***sufficient in the past***, the implications of automated driving on functional safety is set to change that. The absence of the driver availability in the case of unanticipated faults puts ***greater requirements*** on the rigour of design while simultaneously greatly increasing the scope where ***greater rigour*** is required.

Some of the design decisions are related to the core of automated driving such as the decision of the acceptable levels of braking or the acceptable sensor configurations in specific situations. Other decisions may be auxiliary to functionality on, (a) how to assure that the tyre threads have sufficient depth before allowing automated driving to be enabled, or (b) if an ECU not involved in automated driving should be redesigned to be restarted if it starts consuming too much power during operation. Both are examples of decisions that are relevant to automated driving in an indirect way as they could ***indirectly influence*** *functionality* of the ADI because of a change in dynamics (a) or in power available to the ADI (b). A larger discussion on the reuse of legacy Elements in the context of Driving Automation is found in Paper B. Decisions may also be made based on the ***availability of technology*** e.g. by selecting a scanning LiDAR with moving mechanical parts (compromising on a lifetime requirement), instead of waiting for the technological maturity of the alternative Flash LiDARs. A component could be selected based on current availability, even if it does not fulfil a requirement of the ADI, by ***reducing the scope of the function***. E.g. it is easier to achieve automated driving in a particular scenario if the maximum speed is limited, if the route is fixed, or if it is only allowed to drive under clear weather conditions.

The increase in the complexity of the platform with the ADI, and the fast pace of technology progress, also means that design decisions need to be more ***frequently revisited*** during the lifecycle of the platform, and assumptions or choices made in the past could be invalidated based on new information. E.g. It could be found in testing that, due to the positioning of its placement, a camera may be affected by water sprays in certain markets, rendering it ineffective 40% of the time during operation. It is clear that this causes the functionality of the ADI to degrade, but what may be less obvious is that this also affects the safety argumentation.

A safety case covers not only the individual work products which are the evidence for compliance but also the argumentation connecting them showing why they are complete and correct [46]. Suppose that the safety argumentation requires the side camera and a radar to be redundantly monitoring a particular area, would this redundancy be truly valid if the camera were unavailable 40% of the time? A solution could be to reposition the camera; this could have ***cascading effects*** on the algorithm used for classification of objects, or on the transient errors seen due to a change in the length of the cabling required to reach the new position. An alternative solution to reposition the camera, could be to install a wiping system with the camera housing, controlled by another ECU. This could lead to larger power consumption, new control logic requirements, another ECU that must fulfil ASIL classified requirements etc. A design decision made in the absence of concrete test results as is frequently done in the industry, ***relies heavily on assumptions*** made by architects.

Assumptions made on the behaviour of the driver or other road users also affect the safety argumentations. E.g. assumptions on how other road users would react if the ADI switched on hazard lights, or how should conflicting hand signals from cyclists should be interpreted, could affect the controllability measure of a hazardous event. Decisions on core components could also be made based on assumptions e.g. assuming that any emergency vehicle will communicate over a V2V channel and broadcast its location, could eliminate the need for using auditory sensors mentioned in Section 2.4.

Unless these design decisions, assumptions and rationale are traced and followed up in a systematic way with methods relevant to the context, there is a chance that it could lead to a larger cost or a lower level of safety. Safety is a system level property, thus focussing on a top down design like the one advocated by ISO 26262 requires making assumptions about the functioning of the underlying elements. A more detailed analysis of the problem of uncertain information in architecting and its sources can be found in Paper C. According to the beliefs of this author, the need for formalization from the perspective of safety and the limited means of the architecting process to provide this information is a fundamental tension between the two perspectives. While it may never be possible to formalize every aspect of the process of architecting, it can indeed be helped by defined and repeatable ways of working that make the tacit assumptions made during the architecting process, visible and reviewable. A methodology for the systematic architecting of automotive platforms is thus, of critical importance.

## 3.2 Research Questions

### *Part A*

The need for to identify challenges for architecting automated vehicles from the perspectives of safety, business, verification, and validation was considered necessary to scope the work. To determine how the ADI can use the legacy platform, to elicit challenges from the perspectives identified, and to explore the tensions between the perspectives, the research question addressed in Paper A was defined.

> **RQ1. What are the implications of the ways that the ADI can be coupled with legacy platforms?**

### *Part B*

ISO 26262 is the de-facto standard for functional safety in the automotive domain and provides a reference lifecycle to identify and mitigate risk in vehicular functions. The standard prescribes a top down approach for development of safety critical systems that, while ideal for development of a new system, poses significant challenges for the traditional automotive industry practice of reusing existing elements for cost efficiency [24]. The PAA is the first architectural work product in the workflow standard and thus, the ideal point to begin exploratory studies about managing architecting under uncertainty, and the reuse of legacy platforms. Section 3.4 of this thesis defines the PAA as view of the governing Functional Safety viewpoint and uses it as a basis to answer the following research questions:

> **RQ2. How can detailed domain specific information about legacy subsystems be extracted from the platform for the purposes of design of the PAA?**

> **RQ3. How can the decision making of architects under uncertain information be managed to achieve the traceability needed for design of safety critical systems?**

RQ2 and RQ3 have been addressed in Paper C with the creation of the process ATRIUM. RQ2 has in addition been addressed in Paper B with the method M1.

## 3.3 Related Work

The work presented in this thesis is inherently multidisciplinary and even with the delimitations set in Section 1.1, several discourses in the academic community are touched upon such as Design Space Exploration, the reuse of legacy systems, Software and Architecting, Safety and Certification, Dependability, Verification and Validation, Autonomic Systems, design sciences etc. As a result, it is an enormous task to coherently summarize each of these discourses to position this work and thus, this work does not presume to claim full coverage of the State of the art.

An attempt is made in this section to instead summarize those parts of the related work that are of specific relevance to the areas highlighted in Figure 1. To this end, Part A gives a summary of architectures for automation that were found to be interesting in that they had either received significant acceptance, judged by citations, or were representative of a larger set of papers. Part B focusses on the subset of papers that were relevant to dealing with uncertainty in the decision-making process from the software architecture community. Other relevant papers that are not from these discourses to this work are cited as reading guides throughout this thesis as and when their topics are considered.

### *Part A: Identification of challenges*

This thesis aims to address the challenges in architecting automated systems as a primary goal. Due to this papers that addressed automated systems as a whole, from a functional perspective were chosen over others that discussed or improved upon some of the components of automated driving; For example, papers that primarily address object recognition and tracking algorithms, or risk aware planners etc., are not included in this section even though they are considered critical components.

Several publications exist that discuss the architectures of automated driving from a functional perspective. Reference architectures exist as generalizations for automated systems such as MAPE-K [47], OODA [3] as well as domain specific architectures that focus on the needs from a particular domain. NASREM [48], is a typical example of the Sense-Plan-Act paradigm which has found application in the robotics domain [49], while the 4D/RCS reference architecture, an example for unmanned ground vehicles can be found in [50]. For a review of automotive specific reference architectures, the reader is referred to the work of Gordon et al. [51] and Behere [52]. Successful and practical instances of automated driving, judged by their application to real-world case studies, can be found in large research projects such as HAVEit [53], OEM initiatives similar to the Bertha vehicle in [13], participants of competitions such as the DARPA challenges like the “Junior” team in [54], other competitions such as the Hyundai autonomous vehicle competitions [55]. Notwithstanding papers such as [56], which propose fail silent architectural measures to eliminate the need for mechanical backup and those like [57], that are aimed at specific subsystems (and not the entire platform), a limitation seen in these is that they are **too** application itself rather than the interaction with the rest of the automotive platform. While a platform can consist of more than just automation, e.g. subsystems that keep sensors cleaned or climate systems that maintain the comfort of the occupants etc., these are not typically considered even though they could potentially affect the operation of the ADI. Another ***limitation is they do not emphasize on the integration of the ADI into a legacy platform***, arguably the more common scenario, and are limited to the achievement of the ADS functionality.

### Part B: Architecting under uncertainty and reuse of legacy

The principles setting the context for the research questions are not themselves new. Rupanov et al. [58], amongst others, advocate the ***early integration of safety into design***. Sierla et al. discuss the use of functional failure propagation to reduce the risk associated with system design early in the lifecycle, in a way similar to Paper C though without any particular safety standard for guidance [59]. The advantages of using early integration of safety were experienced first-hand by the author in [60], where the use of safety analysis helped structure the architectural design decisions .

Architecting methods for design and evaluation of architecture such as SAAM [61] and ATAM [2] are commonly found in literature. Survey papers of this topic can be found in the works of Dobrica and Niemela in [62], and the authors of [63]. Babar et al. in [64] also present and compare some of the most popular methods in the field, where they come to the conclusion that ATAM is the only one of the considered options that gives detailed guidelines to deal with process issues such as reusability of artefacts, identified risks, scenarios etc. In the initial stages of this work, ATAM was pursued as an option and an attempt was made in the analysis of an automated function with the ATAM attributes augmented with the additional attribute of safety. Though there have been successful cases of the use of ATAM within the automotive context e.g. in Wallin et al. [65] where a method using ATAM is used in decision making for an integration strategy, our attempt unfortunately did not yield successful results. A detailed review of what went wrong with this experiment was not performed but the opinion of the participants was that there was too much overhead compared to the previous process of consensus amongst experts. The DSL used for architectural modelling within Scania was not designed to support documentation in the ABAS style recommended by ATAM. To reverse-engineer a view based on the ABAS style required involvement from a large number of experts, from many different subsystems, which was challenging to organize. Another probable cause could have been the overwhelming nature of the attribute of safety as an inviolable constraint that affected large parts of the platform.

A consequence of the failed application of ATAM was however, a greater understanding of the needs of the methods sought within the ARCHER project. It was found that primary focus needed to be in centred on the early stage design of the architectures for automated driving reusing as much of the legacy systems as possible while not violating the requirements for safety. Also noted was the importance of handling the uncertainty related to the vast changes, that automated driving was due to bring. Of particular concern were the decisions made on whether legacy subsystems were to be reused; and if so, if they needed to be modified to be reusable out of their initial design context.

The study of what constitutes a PAA has been quite limited and fragmented. Searches for the exact phrase “Preliminary Architectural Assumptions” in Google Scholar resulted in only about 40 papers being found. Several papers discussed trends in safety research but only a few discussed the PAA itself. The following is a sampling of papers that discussed the PAA and the characterization of their usage. In terms of definitions, the closest were in [66] where Galina et al. discuss commonalities between ISO 26262, IEC 61508 and ASPICE and Tagliabó et al in [67] who equate the purpose of the PAA to the EAST-ADL Analysis Architecture description. While a few papers discussed the PAA elements in functional terms such as [68], [69]and [70], others such as Taylor et al. in [71] seemingly incorporated hardware elements as well. Westman and Nyberg in [72] use software elements as their building blocks and [73] even include the mechanical considerations such as installation space. A common feature amongst all the papers that were studied is that they did not elaborate how the PAA they used, was generated. Thus, there is seemingly ***no consensus*** within the literature on what the PAA should contain i.e. the definition of the PAA or the methods needed to create it.

The aim of analysing the architecture of a software system is to predict the quality of a system before it has been built; not to establish precise estimates but to examine the principal effects of an architecture [74]. The focus on the level of effects rather than precise estimates or calculations is a reason for the proliferation of heuristics in the field of architectural design. The seminal work of Rechtin and Meier [75] describe the use of ***heuristics as a basic tool*** *in architecting* and works such as [76] have established the effectiveness of using heuristics and the prevalence of such methods in day to day decision making, emphasizing the acceptance of small errors to avoid large mistakes. Maier in [77] discusses the use of heuristics in a progressive fashion from more descriptive to more prescriptive domain-specific heuristics, before the development of quantitative and rational metrics. Heuristic based guidelines have been used in Paper B to satisfactory results.

The need for documentation of not just the requirements of a system but also the ***design decisions-, their assumptions and rationale*** underlying these decisions has been described in [78], where Leveson discusses her construct of Intent Specifications. This concept is already prevalent in other safety critical domains such as that of aircraft certification where ARP 4754A [79] recognizes the traceability between design decisions and safety critical system functionality, as crucial to the design process. Recently however, this has been increasingly observed in the automotive industry as well, with the NHTSA guidelines for automated vehicles also emphasizing the need for ***traceability in design decisions to risks that could impact safety-critical functionality*** [80]. The only two safety reports (that the author is aware of at the time of this publication) released according to the guidelines of the NHTSA from Waymo [14] and by GM [81] discuss the use of risk management through the design stages and claim to adhere to these guidelines. Axelsson in [82] classifies the types of uncertainty in the process of architecting and suggests high-level mitigation strategies for use by architects. In academia, papers like [83] demonstrate the use of rationale with design decisions but the framework they use to manage it is not explicitly shown. Some papers excel at one of rationale management [84], or tracking of assumptions [85], at the expense of the other. Contract based design, proposed for safety critical systems in the automotive industry in papers like [86], are directly benefited by the rigorous tracking of the assumptions made in the design of the system. The need for tracking of decisions and impact analysis of during the iterative development of the system is also noted in [87]. Chen et. al [88] take a similar approach but use design decisions primarily as justification within a safety case. In contrast, the approach taken in this thesis is to focus on the design decisions in a larger context, primarily viewing safety as a constraint rather than the objective of architecting.

Experiences with the attempt to introduce ATAM gave first-hand experience of the overhead of introducing heavy processes into an organization. This overhead may have a counterproductive role in the organization and slow down the development phase if not checked, reflecting the views of Sexton et. al. [89] in that ***balance is needed in the speed of concept design and the assurance of functional safety***.

Thus while none of the principles mentioned so far are new on their own, the combined use of them to create related methods and processes (Paper B and Paper C) appears to our understanding to be a novel contribution. These methods and processes are moreover supported by tools (to support the ATRIUM from Paper C) within the specific context (described in Section 2 and Paper A), make them a methodology for the purposes of Architecting the PAA. The methodology is the main contribution of this thesis.

## 3.4 Viewpoint Definition

This section will define the Functional Safety viewpoint as per ISO 42010 [6] and define the PAA as a view, using terms from Paper B and Paper C. The viewpoint definition in this section was created according to the template and the guidelines suggested by Hilliard in [90].

**Table 3: Functional Safety Viewpoint Definition**

| |
|---|
| ***1. Viewpoint name***<br><br>Functional Safety |
| ***2. Viewpoint overview*** |

This viewpoint is used primarily for the creation and maintenance of the PAA work product, as required by ISO 26262. The PAA view governed by this viewpoint has the goal of documenting rationale of, and assumptions, about design decisions, including but not limited to, on the safe reuse of legacy elements in automated driving scenarios. This viewpoint and the associated view of PAA supports the safety engineering processes by, 1. Explicitly stating the uncertain information used in design decisions for the purposes of impact analysis, 2. Providing architectural input for the purposes of rationale for ASIL decomposition, 3. Assessing and communicating the strengths and the weaknesses of available platform elements, and 4. Justifying the creation of new requirements on existing elements, or the creation of new elements to fulfil functionality, at required integrity levels.

## *3. Concerns and Stakeholders*

### 3.1 Concerns

+ Which architectural alternatives are available for the purposes of ASIL decomposition in the concept phase?
+ What assumptions were made as the basis for a particular design decision?
+ Which variants of a given element have been chosen as acceptable for the purposes of implementation of a particular function?
+ How often has a particular element failed in the field?
+ What are the constraints or dependencies, on the rest of the platform if a particular Design Alternative is chosen?
+ Under which particular configuration(s) of the platform can a minimal risk condition be reached given a particular failure mode according to the current state of analysis?
+ For a particular configuration of a platform, which failure modes have been analysed in the design?
+ Which Assumptions affect a particular element? i.e. Which elements are affected when new information related to the assumption is obtained?
+ Which stakeholder made a particular design decision?
+ Which Design Alternatives were considered in the making of a particular decision?
+ Which Design Alternatives were rejected in the making of a particular design decision and why?
+ Why was a particular Design Alternative chosen over the others?
+ Under which conditions can two Elements fulfil the freedom of interference criterion such that ASIL decomposition can be made?

### 3.2. Typical Stakeholders

The roles of:

1. Architects specifically those who work at the vehicle level.
2. Domain Experts such as ECU designers, function realization engineers, function designers etc.

3. Safety Engineers.

### 3.3. Anti-concerns

This viewpoint is not appropriate or particularly useful to address
- mechanical considerations such as packaging
- refined or detailed technical questions or characteristics

## *4. Model kinds*

The model kinds used in this viewpoint are:

1. PAA_Base
2. Subsystem_Report

## *5. PAA_base*

## *5.1. PAA_base conventions*

### 5.1.1 PAA_base Conventions

#### 5.1.1.1. Model kind languages or notations

A DSL capable of modelling Platform Elements and their interaction from a functional perspective, for the purposes of the meta-model in Figure 11.

#### 5.1.1.2 Model kind meta-model

The meta-model can be found in Figure 11.

## *5.2. Subsystem_Report*

### 5.2.1. Subsystem_Report Conventions

#### 5.2.1.1 Model kind languages or notations

Subsystem report template. A textual report that lists monitors and linked information.

#### 5.2.1.2 Model kind meta-model

Plain text subsystem report that follows the Method M1 described in Paper B, in associating information with specific diagnostic Monitors.

### *6. Operations on views*

**Construction methods:** As described in Paper B and Paper C.

### *7. Correspondence Rules*

Inter-view correspondences:

Sharing of elements with any other views that use the DSL model.

Intra-view correspondences:

Correspondences between Elements of PAA_base and the information from Subsystem_Report.

### *8. Examples <optional>*

### *9. Notes <optional>*

### *10. Sources*

- Specification of the functions in the platform.
- Available FMEAs and Hazard analyses.
- SAE J3016 for functions related to automation.
- Documentation on the DSL used.
- ISO 42010, ISO 26262

# 4. Research design

This section presents an overview of the choice of the research methods used within this work and the motivation for their selection in Section 4.1, before discussing the claims and threats to validity in Sections 4.2 and 4.3 respectively.

## 4.1 Research methodology

A way to visualize the work in this thesis is using the framework described by Checkland and Holwell in [91], where they describe elements essential to any piece of research as a Framework of Ideas, a specific Method and an Area of Concern. Figure 4 is a redrawn [92] image using the elements from the framework, as applied to this thesis. Figure 4 illustrates how, within this framework, ideas of Papers B and Paper C using the method of Engineering Design [93], are applied to the Area Of Concern scoped by Paper A i.e. architecting functionally safe vehicles.

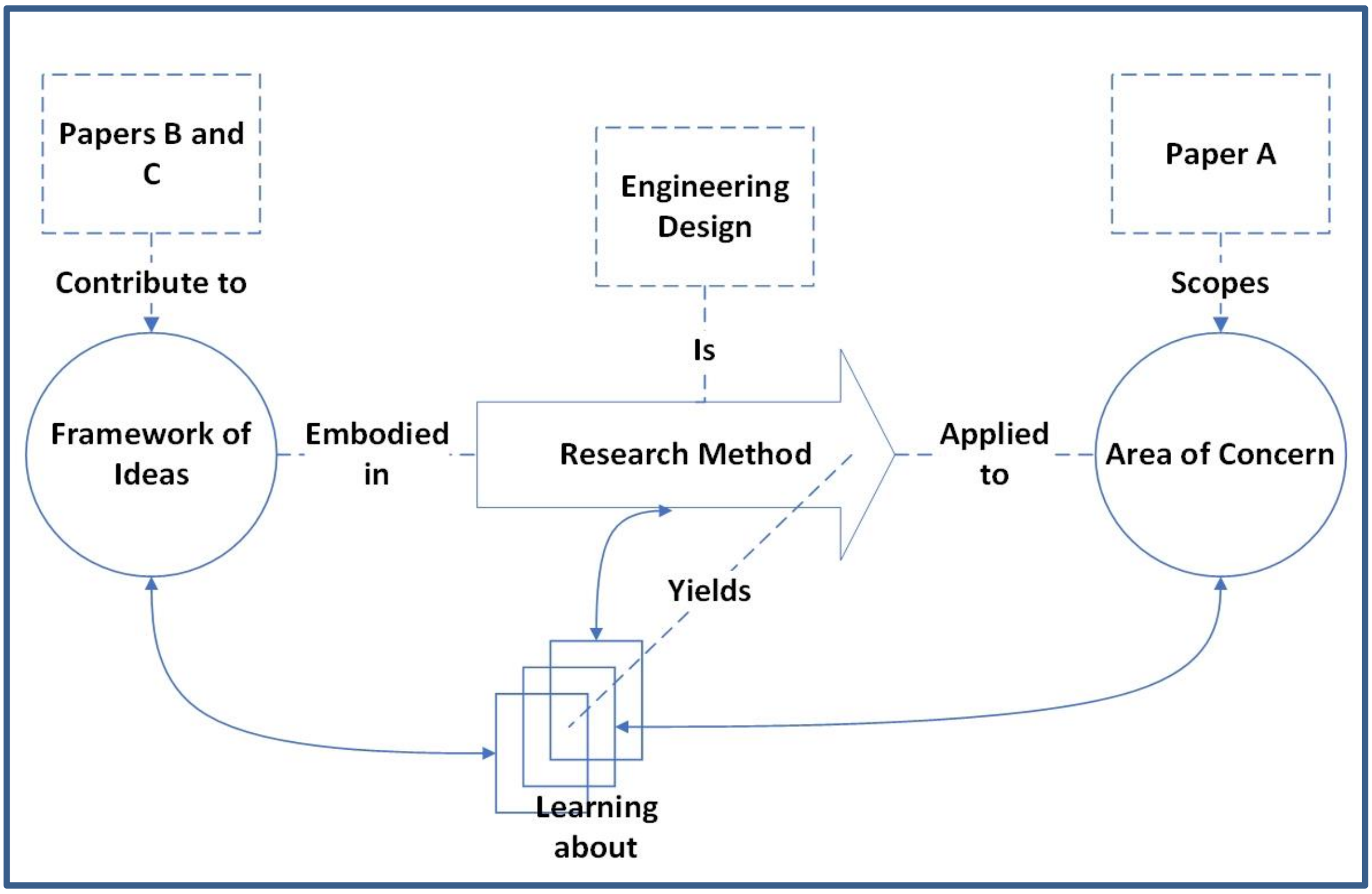

Figure 4: Research Design in Perspective; Adapted from [91]

Research taken as the activity that leads to the creation of new knowledge [94] is, according to Creswell in [95], characterized by the choice of three of its principal components i.e. the choice of the philosophical world-view e.g. constructivist, post-positivist etc., the design of the research e.g. qualitative, quantitative, mixed methods etc., and the choice of specific methods. As seen in Section 3, architecting is related to a system design role, and the architect considers a wide variety of requirements, both functional and extra functional, as inputs and aims to reach a concept that is a satisfactory concept for the stakeholders involved. The multitude of stakeholders, their intertwined concerns, and the need for balancing these concerns to achieve desired qualities, leads to the consequence that research in methodologies for architecting, must necessarily reflect this complex socio-cultural context, to be useful in any fashion. The choice of the pragmatic world-view [95] taken within this research is thus a natural conclusion based on the nature of the domain being studied.

Data collection in pragmatic research also fits very well with the nature of architecting where, depending on the context, the desired abstraction level of the system of interest, and the stage in the lifecycle of the work products, data needed may be available in both quantitative or qualitative forms. Mixed-methods research, as championed by a pragmatic world-view, is not committed to any single system of philosophy, allowing the use of the method that derives suitable knowledge, unconstrained by classifications such as quantitative or qualitative. Data collected and used in this thesis has comprised qualitative and quantitative sources of varying degrees [96]. An overview of the data collected for input and for validation, along with the validation strategies employed, is given in Table 4.

Much has been spoken in literature about suitable research methods in the field of systems engineering. Valerdi and Davidz note the role the results-oriented approach adopted by the current researchers of systems engineering in [97], and also emphasize the need for the right methodology to be selected, as do the authors of [98] and [99]. Müller in [100] gives a range of methods currently in use, discusses the validation challenges of these methods and comes to the conclusion that the current research is limited by the uniqueness of each field situation and making hard claims is close to impossible. These statements apply equally to the field of architecting and make the choice of a specific research method a challenge in itself.

The specific method of Engineering Design as defined by Ferris in [93] aligned with both the goals of the project, and the pragmatic views of the researcher. This thesis as does Ferris, ascribes to knowledge being in of three primary forms of “declarative”, “functional” and “procedural knowledge” as defined by Biggs in [101]. In contrast to natural sciences which targets knowledge of the declarative kind, Engineering Design primarily aims to uncover knowledge of the other two forms i.e. the functional and procedural aspects of *knowing* how the solution to a design problem is formulated functionally, and how it is to be applied. Engineering Design prioritizes the Proof-by-construction method to judge the effectiveness of the design i.e. the building and the analysis of prototypes – which lends naturally to the Case Studies in the papers included.

| Paper | Research Question | Type of Data Source | Data used | Degree of Source Data | Type of analysis | Validation strategies | Degree of Triangulation Data |
|---|---|---|---|---|---|---|---|
| A | RQ 1 | Qualitative | • Identified perspectives and factors of interest in workshop. | First | Qualitative | • Review by industrial and academic participants | First |
| | | Quantitative | • Quantified factors of interest | First | Quantitative, Qualitative | • Review by participants | First |
| B | RQ 2 | Quantitative | • Archival data from industry | Third | Qualitative | • Expert opinion, academic | First |
| | | Qualitative | • Expert knowledge, industrial | First | Qualitative | • Expert opinion, academic | First |
| C | RQ 2<br>RQ 3 | Qualitative | • Requirements analysis<br>• Observations<br>• Expert knowledge, industrial | First | Qualitative | • Longitudinal studies<br>• Workshops<br>• Expert opinion, academic | First, Second |

Table 4: Data Sources in this Thesis

The knowledge developed with the Engineering Design method is *certain* since it solves a problem in it its context, by definition. The certainty of created knowledge is what differentiates Engineering Design from other methods like Action Research [102] where knowledge is considered to be relativist. This differentiation is highlighted in [103] where criteria for a taxonomy of research methods is provided for the ease of analysis. The solution oriented approach of Engineering Design and the principle of finding a satisfactory, distinguished from optimal, product is aligned with the general principles of Design Science Research which has found broad application in the field of Information Systems [104]. The thesis borrows the framework from [105] to ground the engineering design method. Figure 5 redrawn from [106], shows the different cycles in Design Science Research; the Design Cycle is where the artefacts and processes of need are developed, within the context gained from the Relevance Cycle, and where the available knowledge is grounded with the knowledge base of the rigour cycle.

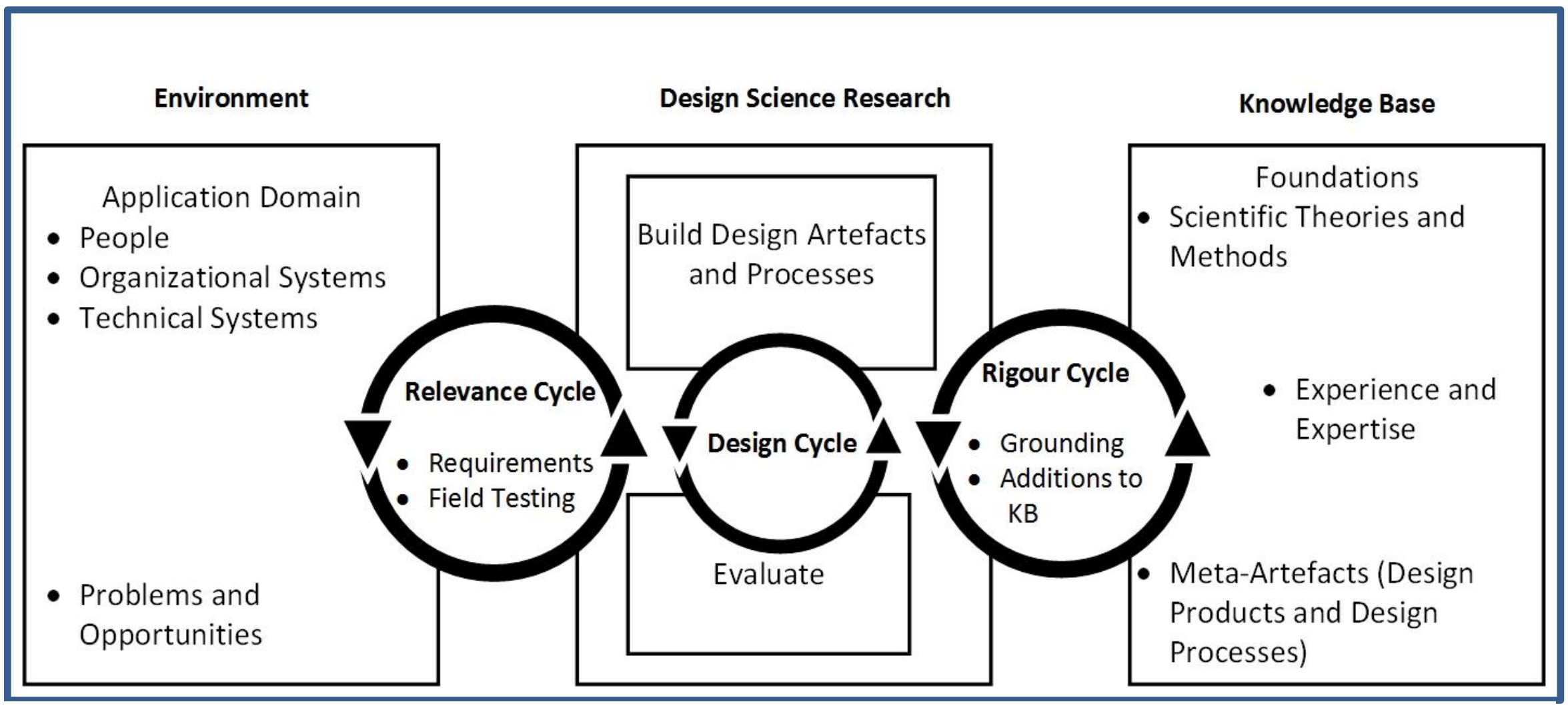

**Figure 5: Research Cycles in Design Science; Redrawn from [106]**

By the scoping provided by Paper A in the relevance cycle, Paper B provides a way to extract information from legacy systems, for use in the process provided in Paper C. The work that has been presented in Section 3.4., Paper B and Paper C, involved several shorter loops in the Design Cycle with the use of toy examples and iterative evaluation by experts. Paper B and Paper C are further grounded in the Rigour Cycle, i.e. experience and expertise of the architect, and industrial best practices such as ISO 26262. They provide design products such as the subsystem reports, and design processes such as the method M1 and the process ATRIUM.

## 4.2 Claims to validity

Robson in [107] gives four different aspects of validity i.e. construct validity, internal validity, external validity and reliability. Using this classification, this subsection briefly describes the claims to validity and finishes by elaborating on the relation to methodology and the threats to validity.

### *Construct Validity*

Construct validity is the measure of how the observed variables reflect the reality of the situation. Two main approaches were used in the included papers to ensure this; the first was by the use of diverse types of studies by using iterative examples, i.e. with progression from controlled toy examples to the realistic studies depicted in the papers B and C, to ensure that the ideas held. The second approach taken was to ensure that both academic and industrial participants were part of the studies. Papers B and C use expert opinions from the industry to construct the methods described therein and utilize reviewers from academia while the exploratory Paper A included a mix of participants from both academia and industry as part of the initial workshops. An advantage of working with industrial case studies is that flaws in construct validity become apparent relatively quickly. Paper B uses heuristics designed and reviewed for ordinary understanding along with detailed examples of application to limit misinterpretations between the researcher's and the users understanding.

### *Reliability*

Reliability relates to the repeatability of the work and the reduction of researcher bias, i.e. will another researcher come to the same conclusions based on the data used. For quantitative methods of analysis, reliability claims are understandably easier to make than for qualitative methods. Thus, while Paper A can make a weak claim for reliability based on the quantification used, Papers B and C which rely on qualitative analysis to a larger degree cannot make such claims. All three papers suffer from this limitation when judged for by reliability by traditional terms. This is a well-known limitation within the naturalistic settings of qualitative research and the method of case studies and has been addressed by papers in literature e.g. by Golshani [108] who advocates the need to redefine these terms for qualitative research. Campbell in [109] takes the view that reliability in qualitative research is achieved through an audit of the procedures used and items such as raw data, data reduction products, and process notes. Following this view, a claim for reliability can be made based on the elucidation of the steps taken in the work, the data and processes used within the thesis.

### *Internal Validity*

Internal validity is the measure of the quality of the findings themselves i.e. if the findings accurately explain the effects observed. The judgement of internal validity has been treated in this work as a function of its utility, as judged by the experts in the review process within the Rigour Cycle from Figure 5. By the utility of the design artefacts created, such as the subsystem report from Paper B, the output deliverables from Paper C and its subsequent comparison to the results from a control group, internal validity can be claimed. A similar claim can be made for Paper A, in that it sets the scope for the other two papers, and the choices made because of its findings, have proven to be useful through the progress of this work.

#### *External Validity*

External validity also known as generalizability or transferability is the aspect of validity that is measured by the extent to which it is possible to generalize the findings. Whether the findings from a work are "generalizable" however, depends on what level the work aims to be generalizable to. This work aspires to be generalizable beyond the primary industrial partner i.e. Scania, within the larger automotive industry. To this end, care was taken during research design to ensure that if there were choices to be made, such as concerns/factors in Paper A, or roles and use of design artefacts such as in Paper B and Paper C, they were general enough to be commonly available across OEMs in the industry or were made on the basis of relevant standards such as ISO 26262. This thesis can claim to be generalizable within this context of automotive OEMs though not necessarily in other domains.

## 4.3 Limitations and Threats to Validity

This subsection expands on the threats to validity (touched upon in section 4.2) by assessing them, and discussing the steps taken towards mitigation, following principles by Runesson and Höst in [110], and Feldt and Magazinius [111].

A long-standing issue that the author has faced in this work has been the barrier, to discuss specific details of the case studies. Instead the papers discuss representative cases and findings. This means that there is a threat that details, about the methodology presented as a contribution in this thesis, may have been inadvertently omitted from inclusion. Unless the methodology is used by other researchers, the threat could lead to different results i.e. a lack of repeatability. This threat has been addressed partly by reviews and changes according to feedback from, presentations at publication venues as well as other industrial and academic workshops, and from engineers that were not part of the work itself, i.e. with no direct vested interest or prior knowledge of the specifics of this work.

Validation of research of a qualitative nature is challenging in general because of the potentially larger influence of subjectivity and biases. The core of the nature of the validity of the work is derived from the utility it provides to the organization using it. The work presented in this thesis is considered to be useful as it is in accordance with standard industry practices, is accepted by the end users, and has been integrated to the larger tool ecosystem within the organization concerned. Though the methodology presented in this thesis has found industrial acceptance, it cannot claim to be the only way to solve the same problem in a similar context. There are only claims of sufficiency through this work, not optimality. Here, the proof is by construction and demonstration such as championed by the method of engineering design.

One of the common themes of this research has been a heavy use of expert opinion, both industrial and academic for the purposes of validation. A threat therefore, lies in the diversity of the experts. For example, though the opinions of various engineers from Scania have been used to shape this thesis, the fact that a large majority of them are employed at the same organization could induce a some bias to the findings. This threat has been considered in for example Paper C where the findings of a control group from a sister organization was used, but was not considered for Papers A and B where the ideas and the analysis itself were more relevant than the actual result (which was assumed to depend heavily on the goals of the organization employing the ideas).

To address the threat of researcher bias, the experts chosen to review the validity have been different from the authors of the papers. Academic experts were chosen to assess the results if the authors have been industrial such as in Paper C, or vice versa such as in Paper B.

Another threat exists in the end-goal of the work i.e. the final quantification step where the field data is to be used to design the PAA and provide a basis for the allocation and decomposition of the ASIL requirements. This thesis has so far not reached the stage where that quantification can take place and given the large number of factors affecting this work, the risk remains that a yet unknown factor may exist that will necessitate changes in the methods of Paper B  and C.

# 5. Summary of Included Papers

This section will discuss what the included papers contribute with individually in Sections 5.1 through 5.3, concluding with a discussion in Section 5.4, on how these contributions tie together to form a methodology for architecting the PAA.

## 5.1 Paper A: Setting the Scope

### *Aim and Research Question*

Paper A is a position paper that addresses the research question RQ1 "**What are the implications of the ways that the ADI can be coupled with legacy platforms?"** by eliciting the challenges for achieving automated driving, from the frame of reference of chosen perspectives.

### *Approach*

The work presented in Paper A was performed in two steps. The first step involved the identification of the different perspectives within the ARCHER project, in a workshop between the stakeholders. The stakeholders comprised of a diverse mix of 4 PhD candidates, 4 senior academics and 3 industrial participants.

The perspectives of interest selected in the first step were

- **Business aspects** i.e. the influences of costs, variability, vehicular platform lifetimes, time to market etc.
- **Functional safety** i.e. the influences of the spread of the ASIL requirements through the platform, the interaction between different safety critical functions etc.
- **Dependability** i.e. decrease in reliability due to the larger number of elements because of both expected failures and higher attack surfaces [112], the need to maintain availability in the presence of this decrease in reliability etc.
- **Verification** i.e. the need for composability of the ADI and its effect on the verification effort needed etc.
- **Realization** i.e. the issues in integrating the cases with the existing legacy platform in a feasible fashion etc.

Having identified the perspectives, 5 architectural cases were chosen representing the diverse ways in which the ADI could be integrated into the legacy Platform. These cases were chosen as appropriate representative cases from across the mechatronic spectrum. Case 1 represents a mechanical way to integrate with the platform. The transition Cases 2 and 3 represent an increasing integration with existing legacy systems without modification. Case 4 represents the case where there is a larger propensity to modify legacy systems to suit the needs of the ADI. The last case is Case 5 where the entire Platform is designed from scratch based on what the ADI requires from the platform. Cases 2 through 4 are represented in Figure 6 along with the symbolic layered Platform where a lower layer indicates a deeper platform integration.

The aim of creating the cases was to gather the concerns of the perspectives influential to

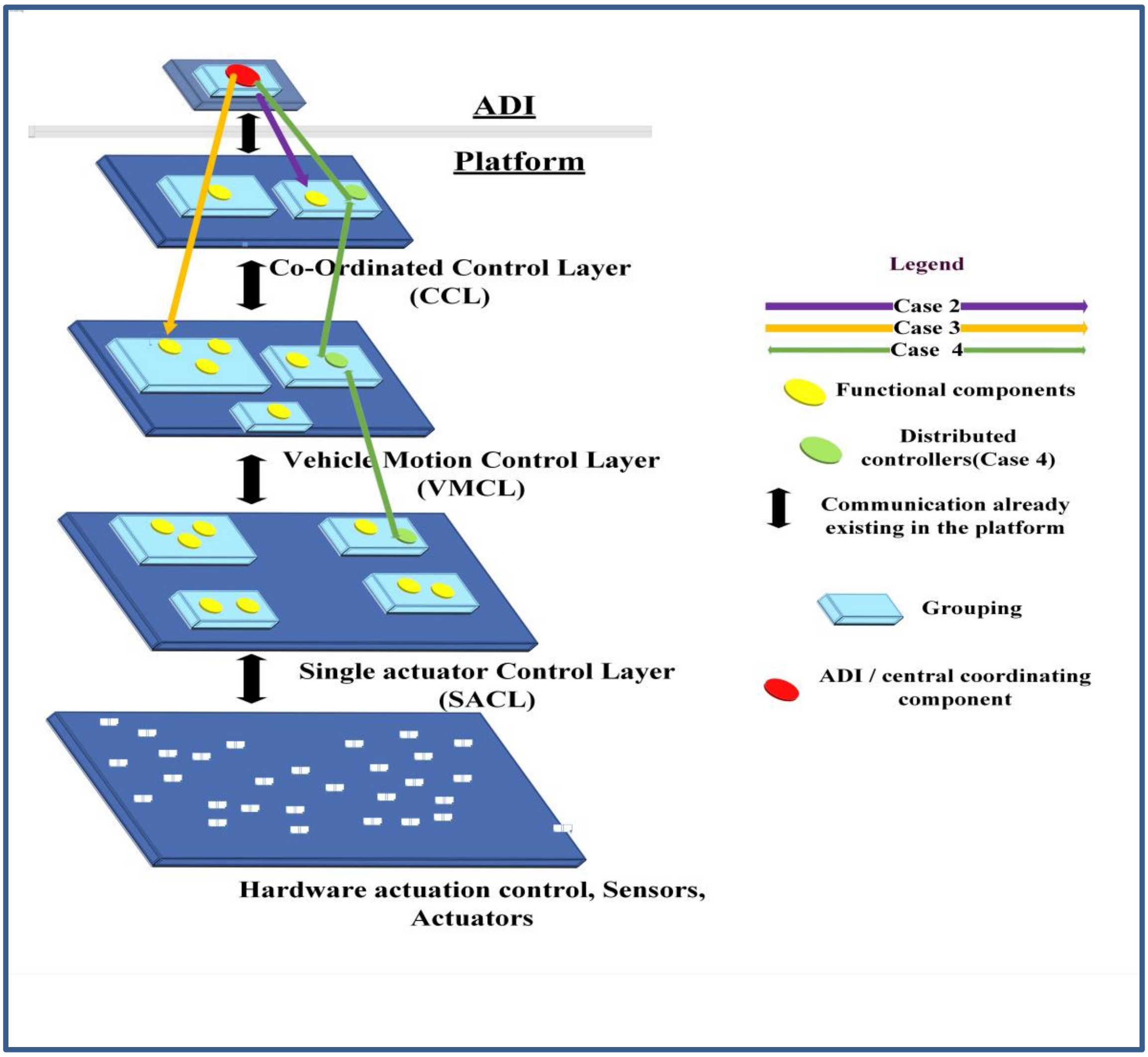

Figure 6: Cases used in Paper A

the task and analysing them in concrete contexts simultaneously, rather than in isolation, thereby trying to highlight the interplay between them. It emerged during the discussion that the perspectives were too broad and that the Cases had various levels of desirability even within a single perspective. E.g. a Case could be rated highly in terms of verification efforts of the platform while being rated low in the verification efforts of the ADI. Factors of interest from within the perspectives where the cases could be rated differently that were identified were:

- Higher Platform Reuse
- Lower accidental Complexity (on reuse)
- Lower Variability (across platform)

- Lower Development Cost Upfront
- Lower Development Cost over time
- Higher Reliability/Availability
- Reduced dependence for Safety Related Diagnostics
- Higher Security
- Ease of Verification of the Platform
- Ease of Verification of the ADI
- Lesser or Controlled Information Flow

In the second step, discussions were held in a second workshop, and the cases were rated according to factors of interest on a scale from 1 to 5 by the participants, leading to the collection of quantified data that was used for analysis.

### Results and Discussion

During the analysis, two different groupings of the cases emerged. The Cases 1-3 were grouped together because of their common characteristic of trying to maintain the legacy systems untouched, Case 3 was the chosen as the most feasible solution in this group. Similarly, Cases 4 and 5 were grouped together by their common characteristic of prioritization of the ADI over the legacy platform, in that the Platform could be modified as needed to fulfil the functional or extra functional requirements of the ADI. Case 4 appeared to be the most feasible for implementation primarily because of the ability for incremental change as per needs rather than a complete redesign. The scoring of the choices from the grouping i.e. Cases

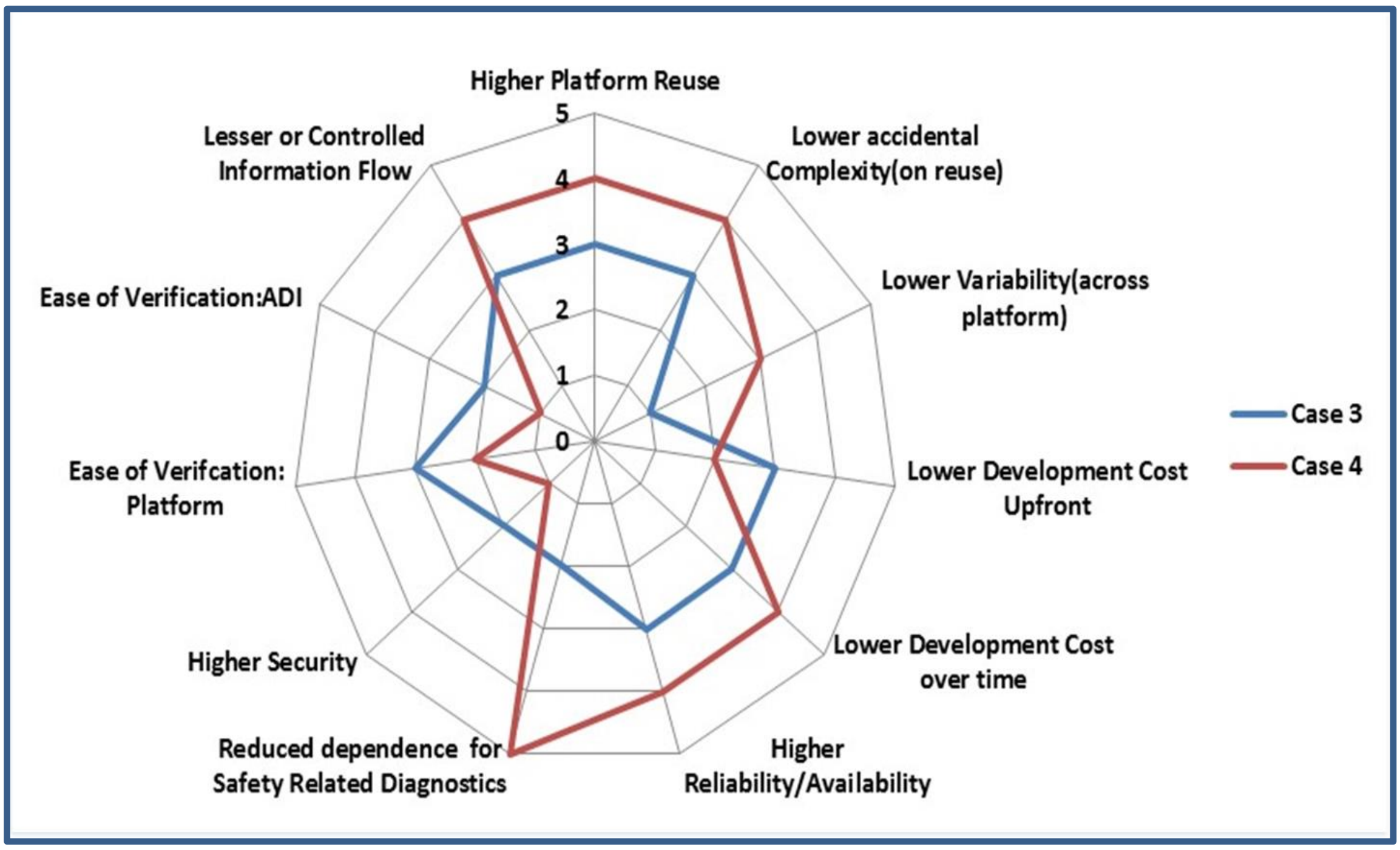

Figure 7: Ratings of the Architectural Cases from Paper A by a Sample Group

3 and 4 are shown in Figure 7, in a radar graph where the distance from the origin is an indicator for desirability, from the point of view of the factors of interest at the edges of the graph. The Cases 3 and 4 were thus chosen to be the most likely scenarios for the addition of ADI, according to the views of the authors.

Paper A thus provided the *main contributions* of, (i) a classification for the diverse ways in which the ADI could be coupled to the platform and, (ii) an analysis from the different perspectives that are of importance to the automotive industry. The contributions address a gap, seen in Section 3.3, in that there is a very limited discourse addressing the addition of automation into legacy platforms. Paper A also served the purpose of scoping the field of interest and enunciating the concerns that would influence both the larger ARCHER project as well as this thesis. One of the most crucial needs from the Business perspective, that of maintaining as much of the legacy platform as possible, received rationale from the other perspectives, and heavily influenced the other papers discussed in the following subsections.

## 5.2 Paper B: Extracting Legacy Information

### *Aim and Research Question*

The need of reusing legacy elements was seen as critically important in Paper A in terms of saving costs, and constraining the design space for the newly designed elements. Section 2.4 discusses the significance of the role of the driver in the design and development of the platform and highlights the importance of reanalysing legacy elements in the new context of automated driving.

Considering that the information of interest was related to legacy systems, i.e. systems whose reuse could be a priority based on extra-functional requirements such as cost savings, a natural conclusion was to try to integrate this information as early as possible in the development process and the safety lifecycle. The PAA was chosen as the point where the information should be integrated.

Paper B attempts to consolidate the needs by focussing on the PAA and by addressing the research question RQ2

**How can detailed domain specific information about legacy subsystems be extracted from the platform for the purposes of design of the PAA?**

### *Approach*

The diversity of the types, and the sheer number, of specialized elements from the platform needed to achieve automated driving meant that the information available to be extracted was equally diverse. The method M1 was designed to address to extract information in a way that would be repeatable across this diversity. In the investigation of how legacy information could be extracted from the platform, the following subtasks needed to be completed.

i. Identify suitable information to be extracted, and its source.
ii. Identify how the information would be extracted.
iii. Identify where and how the result, i.e. the extracted information, would be used within the safety lifecycle.
iv. Identify the level of abstraction needed, and the format for the result.
v. Identify the organizational roles involved and their responsibilities within the method.

The different organizational roles of interest were identified as the roles of the Architects, Domain Experts and the Safety Engineers, with the Domain Experts being the primary users of the method delivering their expertise in a concise form to the Architects. The architects were envisioned as collecting this information from the various legacy subsystems and using this as a basis for the design of the PAA to provide to the safety engineers. With the rationale of reuse, the concept phase from ISO 26262, was seen as suitable point in the lifecycle, for the purpose of early integration of legacy information.

Available work products from completed projects were evaluated to decide the source of the information extraction, and the diagnostic specifications work product was chosen from amongst the available options. The reason for choosing diagnostic specifications as the data source was to obtain a direct connection to the faults and the failure modes of the individual subsystems, since these are relevant to functional safety. The diagnostic specifications comprise information on monitors set in place to detect faults in parts of the subsystem such as conditions for trigger, debounce times, Diagnostic Trouble Codes (DTCs) connected to the monitor etc. The link to the DTCs provides the additional advantage that it allows for a direct connection to field data of occurrence of faults allowing for a potential quantification metric to be added to the selection of a particular set of subsystems. Detailed information on the use of diagnostic specifications in the industry can be found in the ISO standard for specifications for Unified Diagnostic Services [113] or in guidebooks such as [114].

The method M1 is to be used for all subsystems identified by the architect as necessary for the automated driving function in question. Following the method entails for the Domain Expert of a subsystem to follow a set of heuristics in the form of questions and assimilating the answers into a subsystem report. The questions are designed to investigate the effects of the ADI replacing the human driver within the subsystem or in other subsystems. The questions are designed to guide the Domain Expert in finding changes that will be needed to the subsystem in the new context of automated driving. Each question gives a rationale to set the context and examples of concerns and how they could be addressed from the case study. The heuristics thus give a separation of concerns between the roles of the Architect and the Domain Experts. By providing a basic usage context and guidelines to the Domain Expert, the heuristics allow them to abstract their knowledge to a level that is useful to the architects.

The Electronic Braking Subsystem (EBS) used at Scania was used as an object of a case study to judge the utility of the method as the EBS was:

(a) One of the most important subsystems for automated driving; the most capable braking system in the platform
(b) Almost certain to be re-used rather than re-designed from scratch for the first automated driving function

(c) Semi-redundant as there are other methods of braking available in the platform albeit without the same performance levels or robustness. See Figure 8a, for a brief overview of the types of braking systems available in a modern vehicle.

(d) A complex safety critical system that is directly responsible for various basic dynamic and stability related functions e.g. ABS and indirectly responsible for many actuating other functions for safety including those that are legally regulated such as Advanced Emergency Braking systems (AEB). See Figure 8b for an overview of the pneumatic systems controlled by the EBS ECU.

### *Results and Discussion*

The case study looked at about 700+ diagnostic monitors and identified several failures that could be handled by the inherent redundancy present in the platform. It was also possible to isolate about 40 failures of the EBS subsystem that could not be handled within the subsystem. These failures identified aspects that needed to be addressed by either the ADI or a newly designed subsystem upon selection of the EBS for an automated driving function.

The research used quantitative archival data i.e. the diagnostic specifications and qualitative interpretations of it from the Domain Experts from the industry. The Domain Experts and the academic co-authors of the paper judged the utility of the paper based on the case study and the analysis of the method M1 itself. They found it to:

- be useful in concretely aiding PAA design due to its systematic nature of finding requirements for the ADI and the elements of the legacy platform.
- have potential in the quantification of the PAA design by using field data on DTCs to determine failure rates of Elements.

The main contributions of Paper B are as follows: (1) the novel idea of using of diagnostic specifications to facilitate the PAA design (2) the heuristics-based method M1 centred on the removal of the driver for functions related to automated driving.

In the case study the impact of even relatively minor changes to the underlying platform, on the higher layers, became apparent. Safety as a system level property is subject to these intricacies of the subsystems. In choosing to use a legacy system within a new automated driving function, the limitations of the system in this new context, i.e. the absence of a human driver as a fall back, has to be thoroughly analysed so that, (a) new subsystems are designed to cover the limitations within legacy and, (b) there is no overlap of functionality between the new design and the old to save costs. While this is currently done by experienced architects, mistakes are not unheard of. With the advent of automated driving functions and its increase in complexity, there is a concern that there are greater chances for mistakes to be made in the process of architecting. Paper C proposes a way with ATRIUM to formalize the process of architecting within the concept phase, to ameliorate the concern.

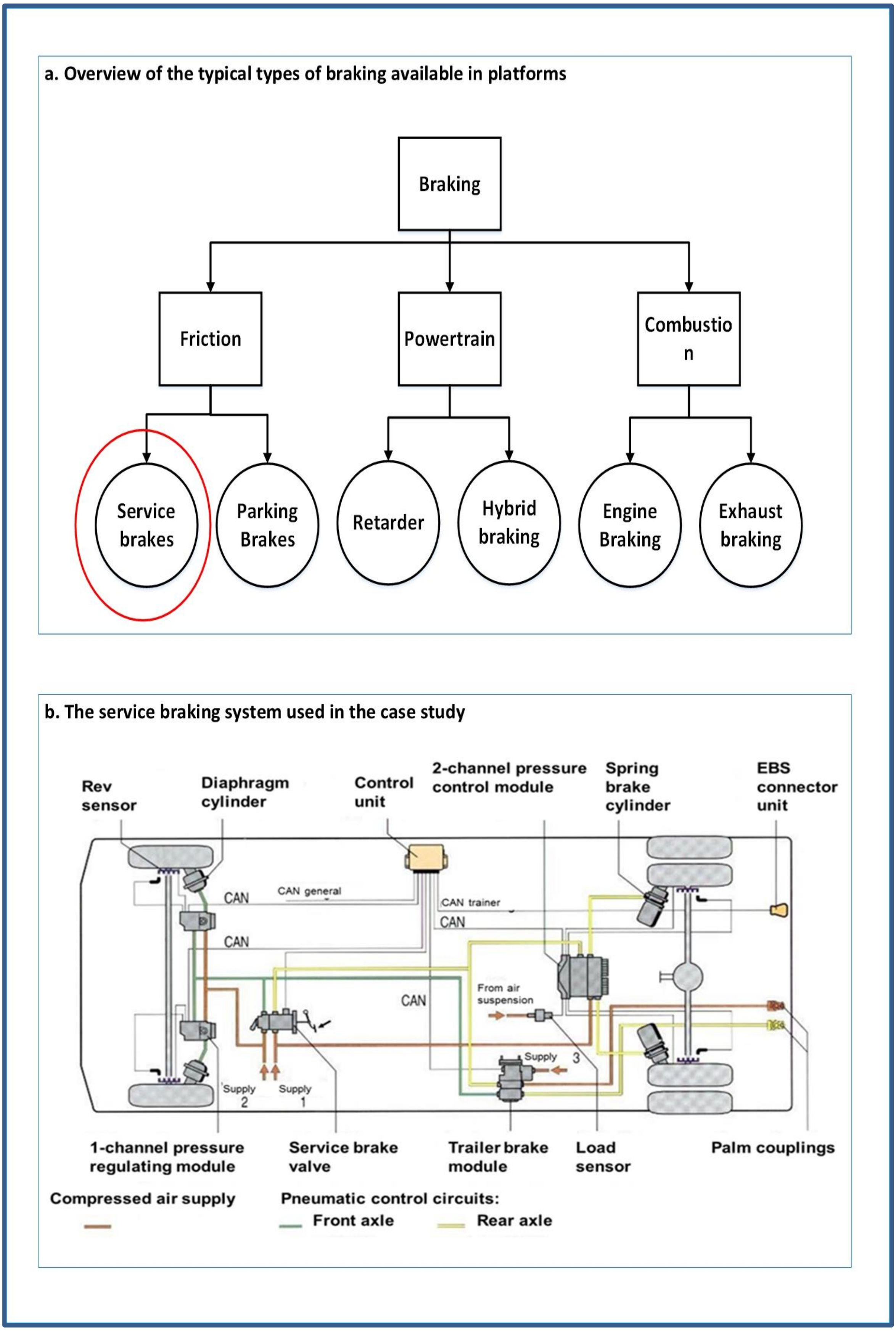

Figure 8: Case Study in Paper B

## 5.3 Paper C: Utilizing Legacy Information & Uncertainty Management

### *Aim and Research Question*

The role of the architect and the uncertainty involved in the process of architecting have been discussed in Section 3.3. But given that the uncertainty is unavoidable in new designs, and in cases of changes in context involving reuse of legacy systems, emphasis needs to be placed to minimize the uncertainty in the process of architecting and to be able to mitigate it where possible.

Paper C thus partially addresses research question RQ2 in defining the use of the legacy information generated from Paper B and primarily addresses the research question RQ3

**How can the decision making of architects under uncertain information be managed to achieve the traceability needed for design of safety critical systems?**

### *Approach*

Paper C proposes a process "ArchiTectural RefInement using Uncertainty Management" (ATRIUM) to mitigate the uncertainty in design decisions and to track the impact of uncertainty in design decisions through the project lifecycle.

The following paragraphs summarize the details of ATRIUM while italicizing the terms introduced by ATRIUM to distinguish them from their nominal meanings in English. ATRIUM provides activities in the form of UML activity diagrams along with a meta-model detailing the relationship between the data elements. An *iteration* represents one complete execution of the process as shown in Figure 9. The core of ATRIUM involves separation of available information into two domains, namely the *Perceived Certain* and the *Uncertain* domain. The information in *Perceived Certain* domain is static and traceability and consistency can be maintained. The *Uncertain* domain contains information that could be subject to change.

*Assumptions* serve as the meeting point between these two domains and transform *uncertain* information into *Perceived Certain* information. An *assumption* may correspond to a functional goal, constraint, operational condition, or any information needed for architectural decisions. The nature of *assumption* is deliberately kept generic to accommodate for the wide variety of inputs that are used by architects. An *assumption* has a single qualifier of *validity*, which can take values of *valid* or *invalid*. *Assumptions* can be added at any point during the process flow. In ATRIUM, Architects primarily work in the *Perceived Certain* domain while the other stakeholders i.e. the experts and the safety engineers work in the *Uncertain* domain. The experts provide the service of clarifying and correcting information in the *Uncertain* domain for the architects.

An *element* has a qualifier called *state*. *State* can take values of *legacy* or *new development*. *Legacy* is assigned as value for the *element* if it has been available before the first *iteration* of the process is initiated. If the *element* is created during the execution of ATRIUM, *state* is assigned the value of *new*. The platform thus is made up of legacy *elements* and new *elements* are added to it with ATRIUM.

A Component Failure Alternative or CFA is a unique combination of a failure mode and an *element*. Each CFA has a single qualifier *state*, which can have either of the two values, *processed* or *unprocessed*. A *processed* value indicates that the CFA in question has been analysed to the best of the current knowledge available within the process *iteration*. An *unprocessed* CFA indicates that either the CFA was never analysed or that additional information at least partially invalidated the CFA analysis.

The inputs referred to by the "gather inputs" abstraction in turn refer to the output of the previous *iteration* (if available), the technology roadmap of the organization and information about the elements.

A *Design Goal* or *DG* is the intended behaviour that the vehicle should achieve in case of a failure. Each CFA is linked to exactly one DG. A DG is comprised of one or more Sub Design Goals (SDGs) which are combined in definite ways to achieve the particular DG using e.g. a time based or a state-based representation. Any SDG might be further broken down into more SDGs as needed i.e. the abstraction level of the SDG is left to the choice of the architects. A Design Alternative or DA is a possible architectural solution that fulfils the DG. Each CFA is analysed separately and if the existing elements cannot fulfil the chosen DG, one or more DAs that enable it to do so are assigned to the CFA. A *processed* CFA can be linked to zero DAs only if elements under consideration already fulfil the DGs. The abstraction of *"Define process parameters"* in Figure 9 refers to deciding the *elements*, failure modes, DGs and generating the CFAs.

*Selection* as a noun is the subset of DAs, out of the set of all possible DAs, chosen to be included as part of the refined PA. A *selection* is made as part of the concluding activity of the ATRIUM *iteration* i.e. in the "generate revised architecture" abstraction in Figure 9.

A *link* is the term used to describe a connection made between informational entities for the purposes of traceability of information. *Links* found in the *Perceived Certain* domain correspond to connections between *assumptions* to CFAs, CFAs to DAs and DAs to the *selection*. Thus, the *selection* can always be back-traced to *assumptions* at any point in the process.

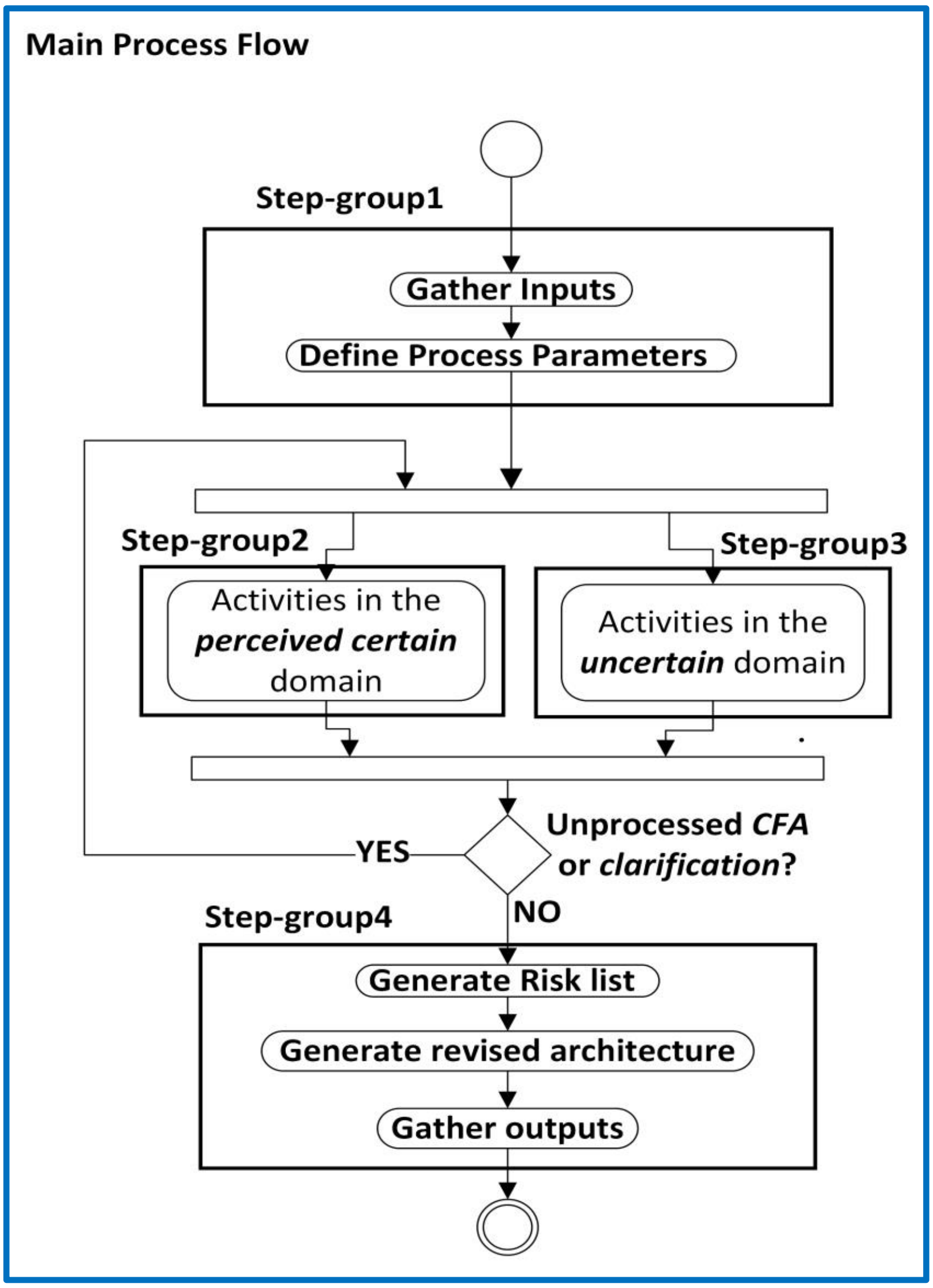

Figure 9: Main Process Flow in ATRIUM

Figure 10b shows the activities of the process in the *Perceived Certain* domain and the management of CFAs. An analysis is performed per CFA and results in zero or more DAs while *assumptions* and *clarifications* are documented along with the appropriate *links*. When the selection is made, the reasons for rejecting the other DAs are recorded along with the reason for choosing the selection.

The abstraction of "process new element" refers to adding new *elements* and populating of the CFAs of those *elements*. The abstraction of "process new assumption" refers to not only adding the assumption to the list, but also the action of reviewing of all CFAs processed so far to see if the assumption has necessitated a new analysis of any CFAs. When new information is obtained that changes or invalidates an existing *assumption*, all linked CFAs are marked as *unprocessed* for subsequent analysis.

Figure 10a shows the process flow in the *Uncertain* domain. *Links* are also found in the *Uncertain* domain and are established from a *clarification* to an *assumption*, or a *task* to an *assumption*. A *clarification* contains detailed questions that must be answered by the secondary users like the experts. All necessary, but uncertain information become *clarifications* and are tracked individually. Each *clarification* requires the creation of an *assumption*, made based on information available and the judgement of the architects to allow work from the architects to progress in the *Perceived Certain* domain. If even the expert does not have access to the information immediately, the *clarification* becomes a *task*. The abstraction *"convert clarification to task"* in refers to this process. A *task* has resources allocated to it and is under the expert's responsibility to complete by a mutually agreed date. The expected dates for completion, name of architect responsible and expert are all documented with the *task*. Unfinished *tasks* at the end of an iteration of the process become *risks* and are collected in the risk list deliverable. This is depicted by the abstraction of "generate risk list" in Figure 9.

A *clarification* and an *assumption* are never deleted. *Clarifications* or *tasks* (when complete) are marked as resolved only after expert consultation (i.e. if the *assumption* linked to the *clarification* was correct or if the existing *assumption* is marked as *invalid* and a new *assumption* is added and linked to both the relevant CFAs and the resolved *clarification*). The link to the piece of information used for a particular *selection* is thus never lost. ***ISO 26262 change management process*** which requires a change request, an analysis, documentation of decision, rationale and implemented changes is thus fulfilled at this abstraction level.

The process *iteration* is completed when (i) there are no *clarifications* remaining, (ii) there are no *unprocessed* CFAs and, (iii) a *selection* has been made. A *selection* is chosen as part of the "Generate Revised Architecture" abstraction. The reason for allowing the *selection* to be made only at the end of the iteration is because DAs may potentially satisfy multiple CFAs, thus limiting the number of changes needed for the system as a whole. ATRIUM does not prescribe any particular method for the selection and allows organizations to use their own.

### *Results and Discussion*

The main contributions of Paper C are the process ATRIUM which gives a flexible way to create and refine the PAA, the analysis of the problem and the discussion of the findings from the case studies conducted.

The linking between the data elements of ATRIUM is shown in Figure 11. The deliverables of ATRIUM are the PAA comprised of the refined Preliminary Architecture (PA) i.e. the *Selection*, the assumption list and the risk list. The way to interpret the results of ATRIUM is that the refined PA is valid under the assumptions listed in the assumption list and is subject to risks documented in the risk list. ATRIUM, in addition to providing a PAA also documents links between failures and assumptions, thus providing easy access to rationales from previous iterations. This enables a basis for consistent discussion, and relevant decisions can be made accordingly and with documented justification.

The requirements based on which ATRIUM was designed were collected after multiple workshops with stakeholders and the process was refined over a few toy examples. Once the process was found to be satisfactory, it was used in the design of an L3 (Level 3 automated driving function) function at Scania to harden a legacy system against single point failures.

The L3 function's known requirements were expanded using the adaptIVe scheme [115] according to the input of the domain experts. The DGs used in the case studies related to ***achieving the minimal risk conditions*** defined for particular CFAs. Several types of minimal risk conditions were identified as needed e.g. based on the physical characteristics of the motion of the vehicle such as the speed and its relative position on the road at the time of the fault detection. Other characteristics of the environment, the ability of the driver to take over, the ability of the system to degrade etc., were also taken into consideration in the selection of suitable minimal risk conditions. Where the behaviours needed from the different DGs had common characteristics, SDGs were defined encapsulating the common behaviours and re-defining the DGs as a sequence of particular SDGs. Additional details about the case study can be found in the appended paper.

The process uncovered several tacit assumptions that were made about the elements (including legacy, new and the ADI), and was useful to ground the requirements for the newly designed elements based on the limitations from the legacy elements. The deliverables were judged to be of better quality than those produced by the previous methods and those produced by a control group of architects working for the same function.

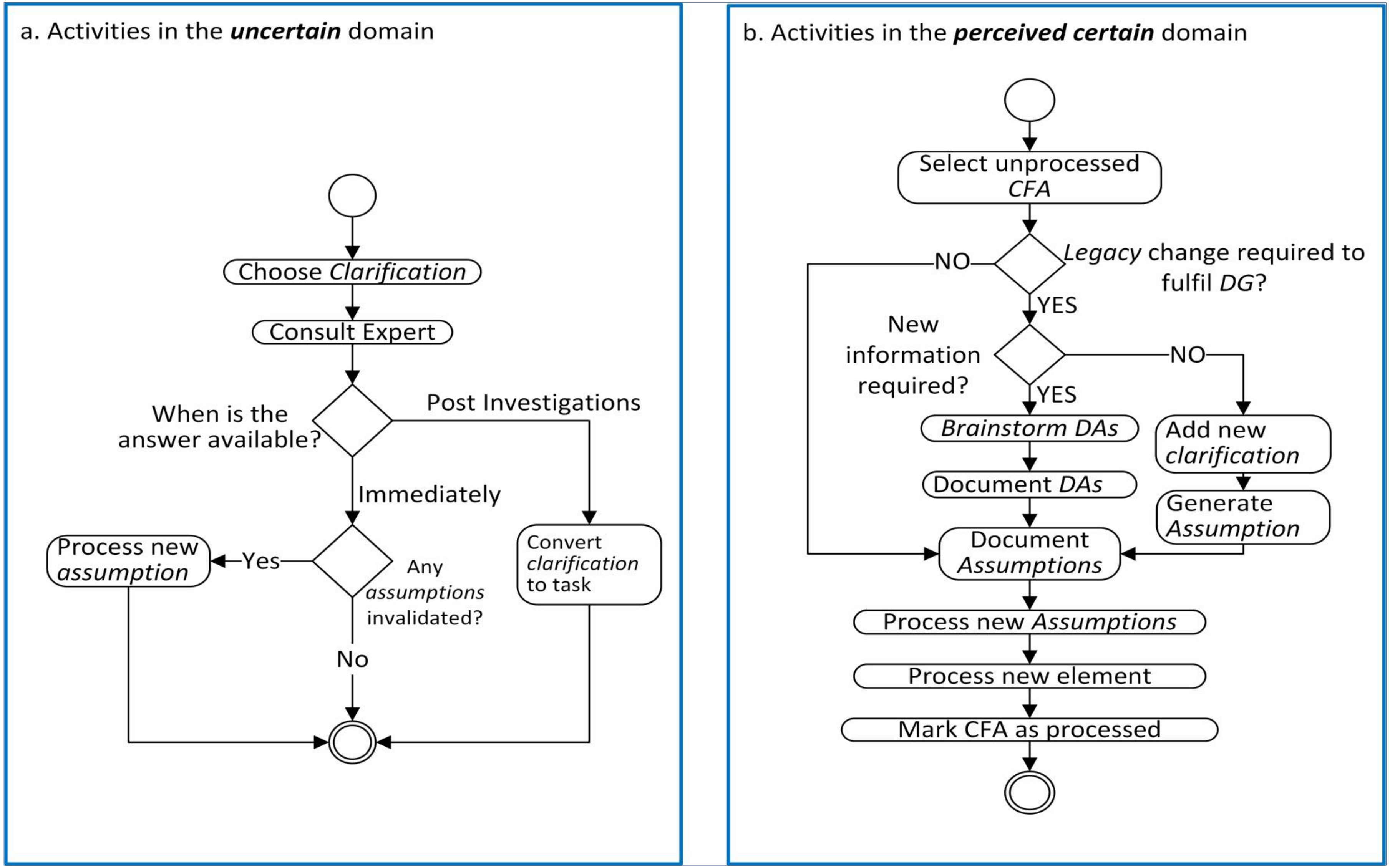

**Figure 10: Activities in the Uncertain and Perceived Certain Domains**

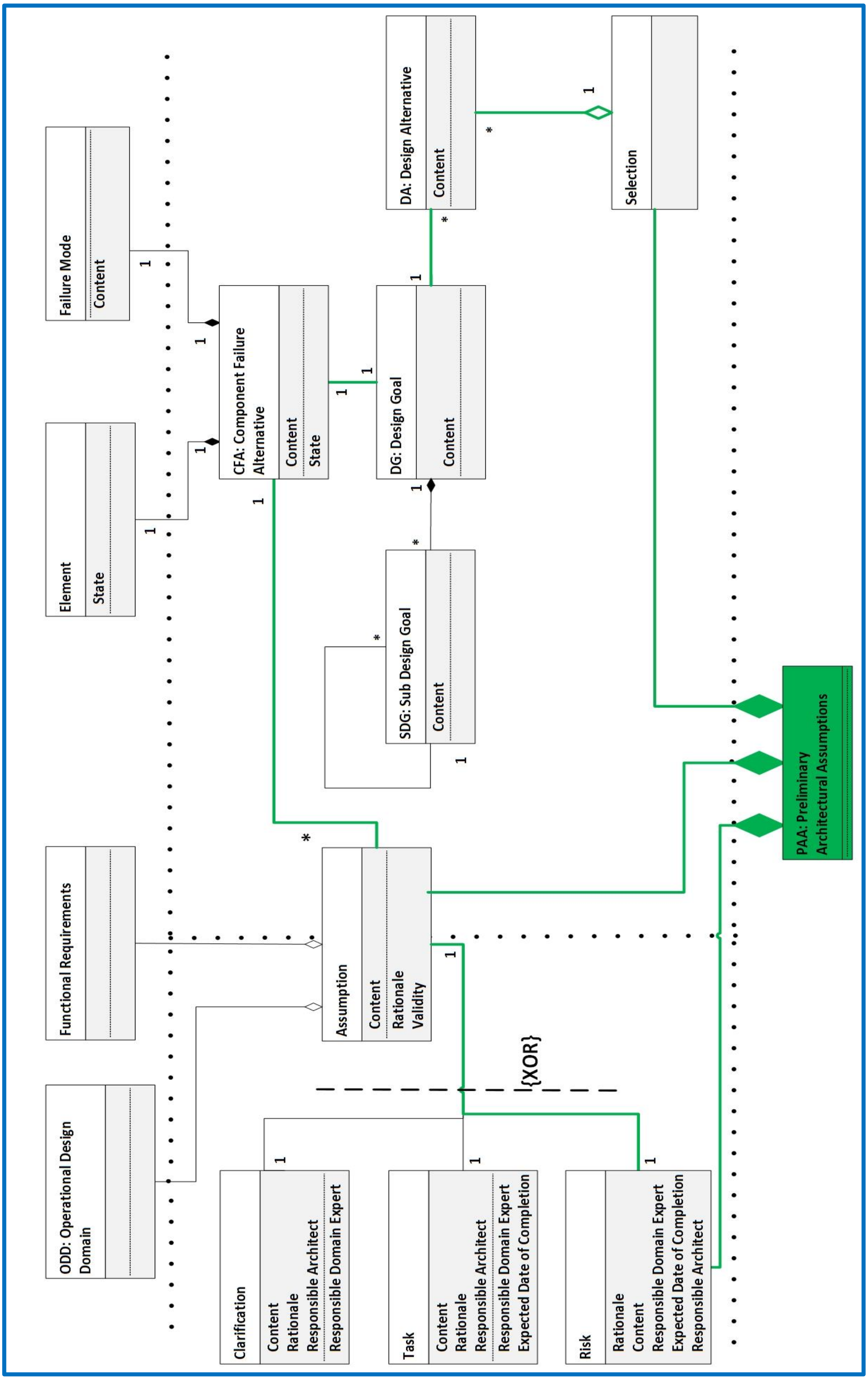

Figure 11: Data Model

## 5.4 Discussion of the Results

This thesis provides a set of related methods, processes and tools (Paper B and Paper C), i.e. a methodology, to achieve the goal of designing the PAA view of the architecture, as defined by the viewpoint from Section 3.4, in the context set by Paper A.

This thesis thus uses automated driving as a context to

- determine the implications to automotive platforms: Paper A.
- motivate the need for reuse of legacy information: Paper A and Paper B.
- design a method to extract legacy information: Paper B.
- position where the extracted information can be used, i.e. the PAA, defining its structure and content: Section 3.4
- reduce the uncertainty in architecting, by grounding the design of new elements on legacy information where possible, i.e. constraining the available design choices: Paper B and Paper C.
- mitigate the impact of the remaining uncertainty by tracking the influence of it over the elements used in the design: Paper C.
- design a process for reusing legacy information and systematically architecting the PAA: Paper C

Legacy based design, as a mid-way solution between the top down and bottom up approaches to architecting, serves to constrain the choices that need to be made in the design of new functions for the platform by leveraging the knowledge and expertise available within the organization. It further helps to reduce the number of iterations needed by a pure top down approach to architecting. Paper A investigates several architectural cases which differ in how the ADI can integrate into legacy platforms and explores the interdependencies between some of the most important concerns in the industry. Allowing only for a small set of architectural cases focussed the scope of the arguments and allowed for a clearer discussion of the concerns faced by each of the perspectives. Findings from Paper A, e.g. the relative futility of trying to keep the ADI strictly separate from the rest of the platform from the safety perspective, the need for the modularization of the ADI from the verification perspective, the emphasis on early delivery and the reuse of legacy from the business perspective etc., helped answer RQ1 and were instrumental in shaping the rest of the thesis.

The aim of the thesis is to help architects in the concept phase to develop safety critical systems. To do so, there was a need to consider the recommendations from the predominant functional safety standard ISO 26262. In the examination of the standard, and the literature concerning the main architectural input in the concept phase, i.e. the PAA, there were no satisfactory definitions to be found. To address this gap in knowledge, the viewpoint described in Section 3.4 was created based on requirements from the architecting process using grounding from the standard, the literature on architecting and expert opinions from senior architects. The viewpoint allows for the definition of the PAA as a view of the architecture according to the viewpoint thereby providing a common definition that can be reused by other OEMs or academia.

For the purposes of defining the problem faced by architects during PA design, the entire platform is envisioned to be divided into three different divisions (i) the functional elements under consideration, (ii) the other elements in the platform and the (iii) ADI. Using this model, the RQ2 can be broken down into the following:

a) Given the selection of a particular functional element i.e. (i), what are the restrictions placed on the selection of others i.e. (ii) and the ADI?
b) Given the answer to a), what new elements must be designed along with the ADI.

The overall conceptual view of the methodology of extracting legacy information and using it as a basis for design of the PAA is shown in Figure 12. Paper B exploits this break-down of tasks to answer RQ2 by delivering a method for the reuse of legacy information to populate the PAA view. Paper B does so effectively analysing the impact of automation at a functional level per subsystem, highlighting the shortcomings and risks of using a particular subsystem with the ADI. M1 utilizes the diagnostic specification to analyse the effects of the absence of human on the vehicle as a whole and enables a domain expert to utilize his exper-tise by identifying potential issues and solutions to these. By using the diagnostic specifica-tions, it brings a gateway to the eventual quantification of the failure rates of the architecture due to the intrinsic connection of the specifications to error reporting in the field. The heu-ristics-based method and the idea of using diagnostic specifications to close the loop between design and operation is seen as contribution.

The knowledge gained by the application of Paper B, the viewpoint from Section 3.4. and guidance from processes within ISO 26262, are then used to strengthen the process of architecting itself in Paper C, thereby answering RQ3. While the methodology does not solve the presence of uncertainty in design decisions, it does make the uncertainty explicit, and opens up the process of architecting for review by making the design decisions and the associated assumptions visible. ATRIUM as shown in Paper C has found application in the industry in several case studies and is the final creation method for the viewpoint from Section 3.4.

The combination of the method, process, supporting tool and the viewpoint gives a powerful methodology for meeting the challenges faced in the architecting of automated driving, and provides an underpinning *for future quantification* of the PAA based on failure rates from the field. Providing a better solution than the current processes in the industry in terms of traceability, management of design decisions, assumptions and rationale, it provides a needed work product that justifies the extra overhead caused by the methodology itself.

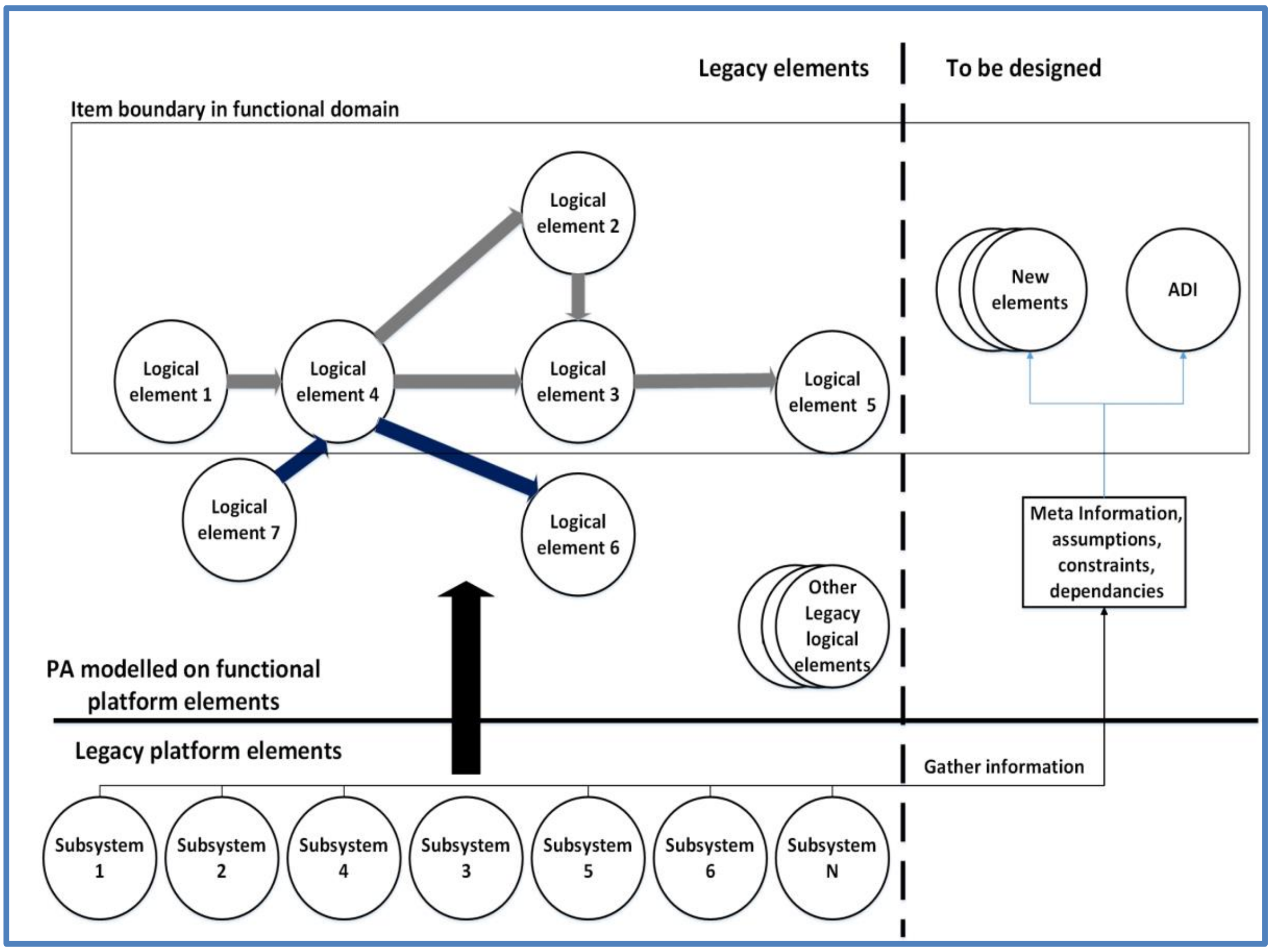

Figure 12: Conceptual View

# 6. Concluding Remarks

## 6.1 Contributions and Impact

The overall aim of this work has been to meet the sub-goal of the ARCHER project in creating methods, tools, and processes to aid architecting, in the face of substantial changes that safe automated driving is expected to bring to vehicular platforms. The sub-goal is considered to have been met in the creation of a methodology for architecting the PAA described in this thesis. A clear definition of the PAA and a way to create it has been missing in the academic work reviewed and in the ISO 26262 itself. The work thus addresses a gap in the safety-critical design of systems and is intended to be a reference for OEMs who wish to improve their processes towards ISO26262 compliance.

The work presented in this thesis has been disseminated at multiple outlets within Scania, publication venues, academic and industrial workshops and has been modified according to feedback received. The presented methodology has been validated by application at Scania both before and after the publication of Paper B and Paper C. Since the publication of the papers, other than further application of the methodology, tool support has been added to aid the data management and to reduce the manual overhead in working with the methods. A prototype of the tool has already been integrated with the Scania tool chain and the methodology is now seen as a standard way of working as it provides the needed role separation and the framework needed for the design of large and complex functions within the platform.

Though not included as part of this thesis, another contribution of the work presented here is the progress made towards the other sub goal of the ARCHER project i.e. the reference architectures for functionally-safe automated driving. During the various case studies undertaken, as the effects of faults in the platform were examined, the number and importance of assumptions made towards run-time risk assessment became evident. This finding was analysed, and solutions were developed to create safety policy-based architectures influenced by supervisory approaches. The solutions include safety policies at design time that constrain the operational envelope of automated driving functions under detected or predicted failures. These policies allow the responsibility of maintaining a predictable operational envelope to be placed on supervisory elements, thereby allowing for the ADI functions to operate freely within the envelope. The solutions are the subject of two patent filings made in late 2017.

## 6.2 Future Work

The work presented in this thesis has strived to facilitate the architecting process for safety-critical, automated, automotive systems. However, this work has been limited in nature. There are further avenues to strengthen the work presented and to extend it. Some of the research directions are as follows,

- Refinement of the methodology from this thesis with continued case studies including:
  - The addition of metrics to facilitate the process of choosing the *Selection* as part of the “Generate Revised Architecture” step in ATRIUM.
  - The incorporation of field data to the methodology to quantify the failure rate of the functional elements.
  - Improvement to the tool designed to aid the methodology
  - Exploratory work in other safety critical domains to generalize the methodology.
- Work on supervisory architectures for functional safety to:
  - Assure an operational envelope that ensures safety while not compromising reliability allowing for a better design-time and run-time split in decision making.
  - Defining the supervisory element’s structure. E.g. its placement within the platform, the need for its distribution etc.
- Validation of architectural concepts by:
  - Classification of critical scenarios for the purposes of automated driving.
  - Creation of a simulation platform where the architectural alternatives can be tested against a defined set of critical scenarios for automated driving.